\documentclass[%
 reprint,
superscriptaddress,
 amsmath,amssymb,
 aps,pre,hyphens,floatfix
]{revtex4-2}

\usepackage[]{amsmath}
\usepackage{mathrsfs}
\usepackage{mathtools} 
\usepackage{enumerate}
\usepackage[usenames,dvipsnames,svgnames,table]{xcolor}
\usepackage{IEEEtrantools}
\usepackage{graphicx}
\usepackage{array}
\usepackage{multirow}
\usepackage{verbatim}
\usepackage{soul}
\usepackage{tabularx}
\usepackage{booktabs,siunitx}
\usepackage[T1]{fontenc}
\usepackage{hyperref}
\usepackage{physics}
\usepackage{siunitx}
\usepackage{makecell}

\newcommand{\mycomment}[1]{}

\newcommand{\av}[1]{\langle {#1} \rangle}
\newcommand{\stk}[1]{\ifmmode\text{\sout{\ensuremath{#1}}}\else\sout{#1}\fi}

\AtBeginDocument{\RenewCommandCopy\qty\SI}

\begin{document}

\title{Dynamical heterogeneity and universality of power-grids}

\author{B\'{a}lint Hartmann}
\email[]{hartmann.balint@ek.hun-ren.hu}
\affiliation{Institute of Energy Security and Environmental Safety, HUN-REN Centre for Energy Research, P.O. Box 49, H-1525 Budapest, Hungary}

\author{G\'{e}za \'{O}dor}
\email[]{odor@mfa.kfki.hu}
\affiliation{Institute of Technical Physics and Materials Science, HUN-REN Centre for Energy Research, P.O. Box 49, H-1525 Budapest, Hungary}

\author{Istv\'an Papp}
\affiliation{Institute of Technical Physics and Materials Science, HUN-REN Centre for Energy Research, P.O. Box 49, H-1525 Budapest, Hungary}

\author{Krist\'of Benedek}
\affiliation{Budapest University of Technology and Economics, Műegyetem rkp. 3, H-1111 Budapest, Hungary}

\author{Shengfeng Deng}
\affiliation{Institute of Technical Physics and Materials Science, HUN-REN Centre for Energy Research, P.O. Box 49, H-1525 Budapest, Hungary}

\author{Jeffrey Kelling}
\affiliation{Faculty of Natural Sciences, Chemnitz University of Technology, \\ Stra{\ss}e der Nationen 62,  09111 Chemnitz, Germany}
\affiliation{Department of Information Services and Computing,
Helmholtz-Zentrum Dresden-Rossendorf, P.O.Box 51 01 19, 01314 Dresden, Germany
}


\begin{abstract}
 While weak, tuned asymmetry can improve, strong heterogeneity destroys synchronization in the electric power system. 
 This study explores  and compares the heterogeneity levels of high voltage (HV) power grids in Europe and North America.
 We provide an analysis of power capacities and loads of various energy sources from the databases and found heavy
 tailed distributions with similar characteristics. Graph topological measures and community structures also 
 exhibit strong similarities, while the cable admittance distributions can be well fitted with the same power-laws (PL),
 related to the length distributions. The community detection analysis shows the level of synchronization in different
 domains of the European HV power grids, by solving a set of swing equations. We provide numerical evidence for 
 frustrated synchronization and Chimera states and point out the relation of topology and level of synchronization
 in the subsystems. We also provide an empirical data analysis of the frequency heterogeneities within the Hungarian HV network 
 and find q-Gaussian distributions related to super-statistics of time-lagged fluctuations, which agree well 
 with former results on the Nordic Grid.

\end{abstract}

\maketitle

\section{Introduction\label{sec:1}}

Electric power systems under disturbances are often studied using variations of the Kuramoto-model. A notable shortcoming of this approach is that it relies on simplified models of the grid under study, including its infrastructural properties and its behavior, which may lead to qualitatively wrong results {\cite{RODRIGUES20161}}. While agreeing that researchers are forced to make simplifications, mainly because open data sources are rare and incomplete, bridging this gap is inevitable to effectively transform theoretical results into practice. For this reason, the present paper focuses on completing the grid data-sets using experience from the domain of power systems. Another goal of the authors is to use this process to deepen the understanding of dynamical heterogeneity and universality in power grids, continuing the work presented in \cite{odor2020}.

Cascade failures in power grids occur when the failure of one component or subsystem causes a chain 
reaction of failures in other components or subsystems, ultimately leading to a widespread blackout or outage
and disintegration of the network~\cite{PhysRevE.69.025103}.
Historical data \cite{carreras2000} and self-organized criticality (SOC) models \cite{SOC} on direct current 
(DC) model simulation ~\cite{dobson2007c} have shown that blackout size distributions, measured by 
various quantities, energy, power, duration, can be described by fairly universal power-law (PL) 
tails~\cite{carreras2004,dobson2007c} and SOC is expected to arise by the competition of 
power-grid supply and demand. To explain the scale-free behavior, relation to the PL of city-size distributions has also been suggested~\cite{nesti2020}.
Various other explanations also exist, for example, giving a special power source 
distributions from the consumer side, that leads to PL-s.

Cascade failures in alternating current (AC) models have also been investigated ~\cite{M1,M2,M3,M4,M5,SWTL18}. 
Symmetry breaking in the network parameters, especially in the phase shifts between neighboring nodes, have been found to stabilize the synchronization~\cite{Molnar_2020},
albeit fine tuning of the asymmetry in the mass generators is essential, without this, local damage can cause fatal blackouts~\cite{10.1063/5.0131931}. 
Our previous studies based on solving the swing equations on different power-grids
have arrived to different conclusions.
While for a full high-middle-low voltage (HV--MV--LV) synthetic network we found 
higher stability than on the 2 dimensional homogeneous lattice of same size~\cite{POWcikk},
in the HV networks of Europe and Hungary, strong heterogeneity drastically reduces 
global synchronization~\cite{Powfailcikk}. 
An important, challenging question that remains is the determination of network topological effects
and regions of enhanced stability, which can be obtained by symmetry breaking \cite{PhysRevE.71.016116}. 

Very recently, PL blackout size distributions have also been confirmed via AC modeling,
using the numerical solution of the swing equations near the vicinity of the 
power-grid synchronization point~\cite{POWcikk,Powfailcikk,USAEUPowcikk}.
Here generally Gaussian distributed self-frequencies have been used.
One can ask, whether the synchronization stability or the properties of cascade failures are altered by various heterogeneities (i.e. exponential or PL distribution of consumer and generator powers). 
Earlier we found that these kind of modifications in solutions of the swing equations lower the synchronization, but did not change the forms of failure cascade size distributions 
or the range of their occurrence in the control parameter space~\cite{Powfailcikk} .

More generally, to explain PL electrical outage statistics, we proposed other mechanisms 
following the spectral analysis of large outage duration data sets.
According to this, the observed auto-correlations in data at shorter times  
imply a SOC mechanism of the competing maintenance supply and demand, leading 
to branching failure cascades. For larger times, the lack of such
correlations suggested a highly optimized tolerance (HOT) mechanism~\cite{outagecikk}.

In this study we provide a large-scale data analysis of power sources and consumers and investigate transmission 
line admittances and network communities in HV power-grids of Europe and north America. We shall point out
that, contrary to the existing large heterogeneity, main features seem to be universal and may provide 
explanation for the frequent occurrences of power laws in the outage distributions.
The level of universality, of course, is just approximate, as various previous studies have
found it~\cite{Pgridtop,Martins_Ribeiro_Oliveira_Forgerini_2018,POWcikk,Powfailcikk}. 
Our results imply that PL emerges on continent sized networks, while sub-systems exhibit 
deviations from it.
Here we go beyond static and graph theoretical analysis and  provide a dynamical study 
by determining the graph communities and the synchronization in them via solving the 
swing equations. 
Network community analysis is a fundamental tool to explore topological heterogeneities. 
The role of communities, as strongly connected domains, can be very profound and may provide 
rare region effects, altering the dynamics of the system
\cite{Griffiths,Vojta,HMNcikk}. The modular structure can also enhance frustrated
synchronization~\cite{Frus,Frus-noise,FrusB,CCrev} and Chimera states~\cite{chimera,Flycikk}.

While many publications tend to use open-source and/or generative models (such as \href{https://www.power.scigrid.de/pages/downloads.html}{SciGRID}, \href{https://github.com/bdw/GridKit}{GridKit}, \href{https://pypsa.org/}{PyPSA}, \href{https://wiki.openmod-initiative.org/wiki/Transmission_network_datasets}{OpenMod Initiative} or the \href{https://open-power-system-data.org/}{Open Power System Data} models), these typically lack the level of detail necessary for case studies. To reflect on how such situations can be handled, here we provide a way to estimate missing admittances and graph edge weights for the Kuramoto model \cite{kura} upon some assumptions.

\section{Heterogeneity and universality in empirical power-grid data \label{sec:2}}

In this section, a statistical analysis of various empirical data released to European power grids is presented. The non-Gaussian distributions
obtained, underline the necessity to perform numerical modeling by
more detailed models, based on swing equations to be discussed in the following.
Besides, we determine communities of the 2016 EU network by which
the local synchronization analysis, started in Ref.~\cite{USAEUPowcikk} 
is extended.

The time evolution of power-grid synchronization is described by the swing
equations~\cite{swing}, set up for mechanical elements (e.g.~rotors in generators 
and motors) with inertia.
It is formally equivalent to the second-order Kuramoto equation~\cite{fila},
for a network of $N$ oscillators with phases $\theta_i(t)$. 
Here we use a more specific form~\cite{Olmi-Sch-19,POWcikk},
which includes dimensionless electrical parametrization and
approximations for unknown ones:
\begin{equation}\label{kur2eq}
{\ddot{{\theta }}}_{i}+\alpha {\ }{\dot{{\theta}}}_{i}=P_i
+\frac{{P}_{i}^{max}}{{I}_{i}{\ }{\omega }_{G}}{\ }\sum
_{j=1}^{N}{{W}_{\mathit{ij}}{\ }\sin \left({\theta }_{j}-{\theta
}_{i}\right)} \ .
\end{equation}

Here $\alpha$ is the damping parameter, which describes the power dissipation,
or instantaneous feedback~\cite{Powfailcikk}, we define $K:=P_i^{max}$ as a global control parameter, related to the maximum transmitted power between nodes, the inertia $I_i=I$ and the nominal generator frequency $\omega_G$ are kept constant in the
lack of our knowledge; and $W_{ij}$, is the adjacency matrix of the 
network, which contains admittance elements, calculated from impedances to be described 
in Sect.~\ref{sec:3a}.
The quenched external drive, denoted by $P_i:=\omega_i^0$, which is proportional
to the self-frequency of the $i$-th oscillator, 
carrying a dimension of inverse squared time $[1/s^2]$, 
describes the power in/out of a given node and Eq.~(\ref{kur2eq}) 
is just the swing equation (phases without amplitudes) of an AC power circuit.
As commonly done with the first-order Kuramoto model, the self-frequencies 
are drawn from a zero-centered Gaussian random variable as rescaling 
invariance allows to gauge out the mean value in a rotating frame. 
For simplicity, one can assume that $\omega_i(0)$ is drawn
from the same distribution as $\omega_i^0$ and numerically set
$\omega_i(0)=\omega_i^0$. 
In this study, the following parameter settings were used:
the dissipation factor $\alpha$ is chosen to be equal to $0.4$ to
meet expectations for power grids, with the $[1/s]$ inverse time 
physical dimension assumption.

\subsection{Heterogeneity in grid frequency distributions}

\begin{figure}[!htbp]
    \centering
    \includegraphics[width=1\columnwidth]{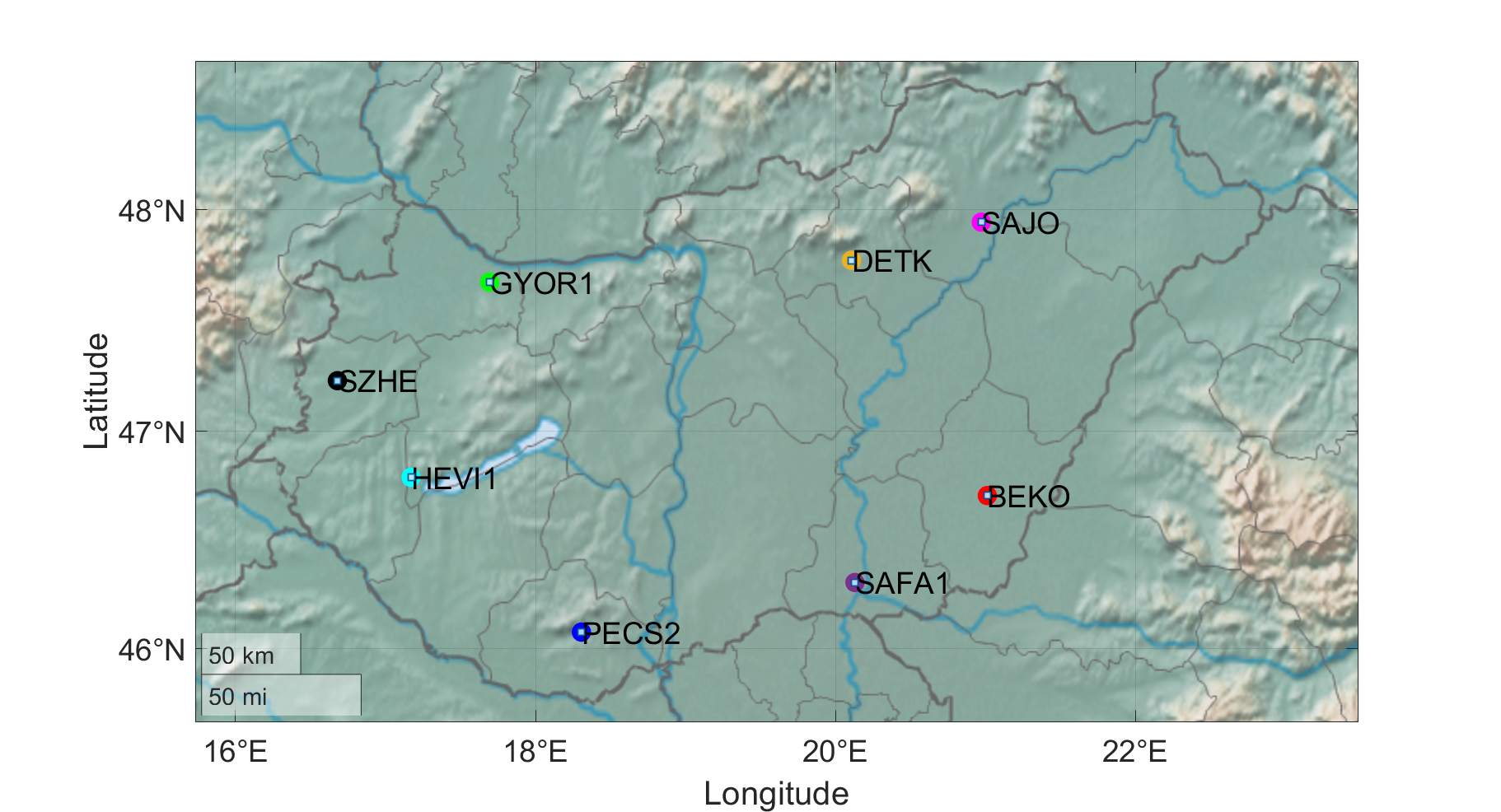}
    \caption{Locations of frequency measurement sites used for the analysis. Abbreviations refer to 400 kV substations, as follows: BEKO - Békéscsaba, DETK - Detk, GYOR1 - Győr, HEVI1 - Hévíz, PECS2 - Pécs, SAFA1 - Sándorfalva, SAJO - Sajószöged, and SZHE - Szombathely.}
    \label{fig:freq_meas}
\end{figure}

\begin{figure}[!htbp]
    \centering
    \includegraphics[width=0.9\columnwidth]{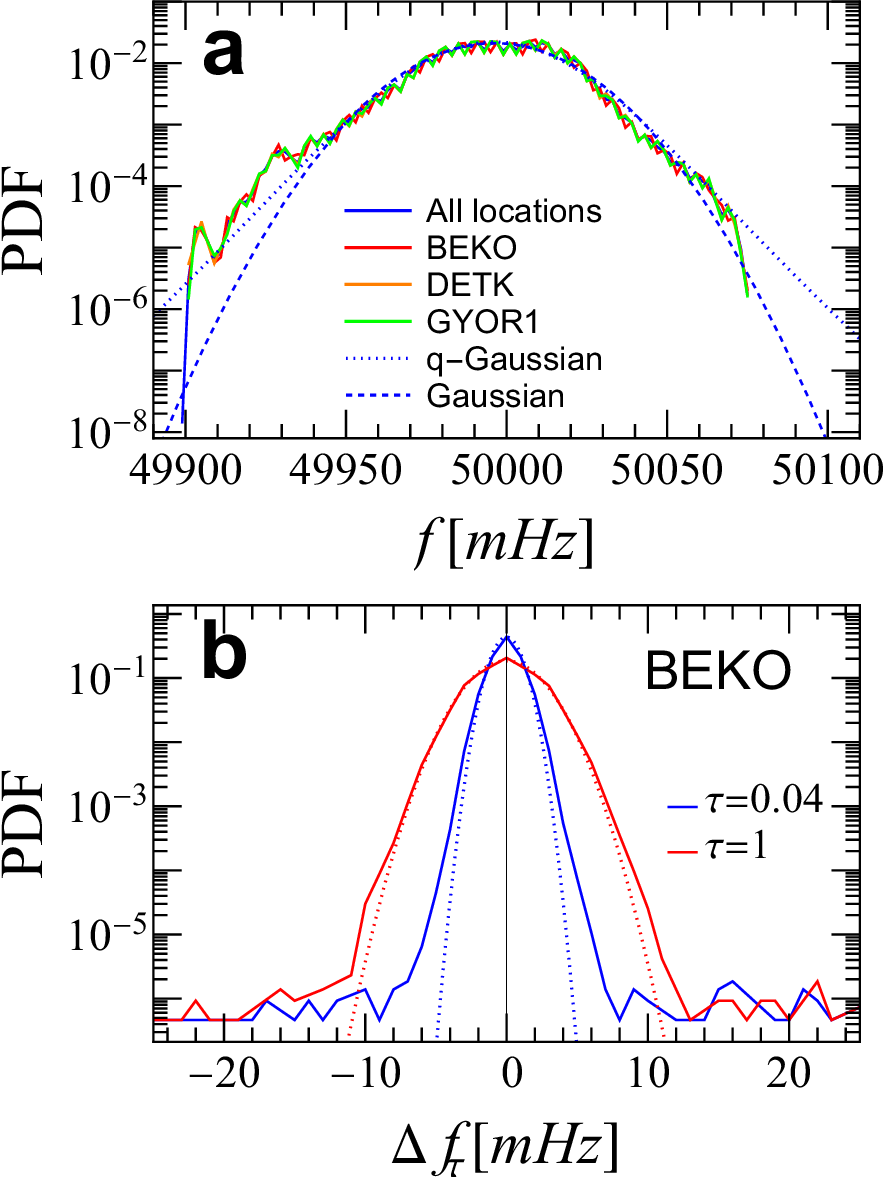}
    \caption{Probability density functions (PDF) of (a) frequencies measured at different locations and of (b) increments of BEKO at time lags $\tau=0.04s$ and $\tau=1s$. In (a), the Gaussian fit is $\mathcal{N}(\num{49996.2601}, \num{18.9679})$ and the $q$-Gaussian fit is given by the expression $\frac{0.0220}{\left(1 + \num{0.000179} (\num{49996.3} -x)^2 \right)^{9.274}}$ with $q\simeq 1.1078$. The dotted lines in (b) are the corresponding Gaussian distribution fits.}
    \label{fig:freqdist}
\end{figure}

From the Kuramoto equations, one sees that the frequencies of the oscillators constitute an important variable to power-grid dynamics, not only because the nominal frequency $\omega_G$ may show slight spatial variation, but also because the measured frequency $\omega(t)$ will fluctuate in the vicinity of $\omega_G$ due to the system dynamics~\cite{schafer2018}, despite the efforts to bring the system to synchronization.

The modeling of frequencies in the Kuramoto equations has been presented with various approaches in the last decade and a half \cite{RODRIGUES20161}, as the research community did not only search for more accurate representations but also examined the effect of frequency distributions on synchronization and stability as well. In their early work, Filatrella et al. \cite{filatrella2008analysis} suggested that a bimodal distribution of frequencies is the most appropriate one for the power grid, a consideration which served as a basis for later studies. A different approach was presented by Olmi et al. \cite{olmi2014hysteretic}, who used natural frequencies randomly distributed according to a Gaussian distribution, and by Pinto and Saa \cite{pinto2016synchrony} and Taher et al. \cite{taher2019enhancing}, who drew initial frequencies from uniform distributions.

The authors of the present paper have shown in~\cite{PhysRevE.98.022305} that the addition of a stochastic noise, modeling random frequencies of distributed generation does not affect the forms of desynchronization distributions. They also highlighted that the results imply that heterogeneous networks have better performances than what homogeneous approximations could predict.
This was based on the comparison of the Kuramoto solution on an $N=10^6$ node synthetic HV-MV-LV power-grid with that of $N=1000^2$ 2D lattices. However, here a small, 32 node, bidirectionally symmetric, looped HV part determined the behavior of the whole MV-LV part, which was attached at MV codes as unidirectional cluster chains of consumer nodes. Thus presumably the crossover size effect causes the stability in this network topology. In their subsequent work \cite{odor2020}, they replaced Gaussian self-frequencies by exponentially distributed ones, which led to a drop in the steady state synchronization averages, but did not affect cascade size distributions.

Recently the focus of research was shifted to the analysis of frequency measurement data and their implication on synchronization. Wolff et al. \cite{wolff2019heterogeneities} found after investigating over 100 histograms that local frequency deviations from the nominal frequency show Gaussian nature for small deviations and an exponentially decaying tail part. They also highlighted that the standard deviations were proportional to the magnitude of power injected at the node by distributed generation.  In their papers \cite{rydin2020open,gorjao2021spatio,anvari2020stochastic}, Rydin et al. present an in-depth analysis of high-resolution frequency measurements. They conclude that histograms are good indicators of how heavy-tailed the frequency distributions are, and that measurements of frequency auto-correlation functions reveal information on important patterns, likely connected to phenomena specific to the examined synchronous area. They also show that synchronously recorded frequency data exhibits very complex spatio-temporal behavior and small fluctuations around the mean follow different types of super-statistics. A subsequent work \cite{oberhofer2023non} capitalizes on these experiences and proposes a Fokker--Planck equation to extend stochastic power-grid frequency models to handle non-Gaussian statistics as well. A different stochastic process, an Ornstein--Uhlenbeck process is suggested by \cite{kraljic2022towards} to model statistical properties (e.g.~double-peaked probability density functions, heavy tails) of frequency measurements.
Finally, Jacquoud et al. \cite{jacquod2022propagation,tyloo2023finite} show that non-Gaussian fluctuations of frequency decay with the distance from the source faster than Gaussian ones do, but such noise also propagates through the whole grid, resulting in voltage angle fluctuations resembling the same non-Gaussian distribution.

To demonstrate the variation and fluctuation of power-grid frequency, we show empirical frequency data measured synchronously at eight different locations across Hungary within 24 hours on Oct.~23 2022~\cite{MavirFdata}. The nominal synchronous frequency of the Hungarian Grid is 50 Hz and the recordings were taken at Békéscsaba (BEKO), Detk (DETK), Győr (GYOR1), Hévíz (HEVI1), Pécs (PECS2), Sándorfalva (SAFA1), Sajószöged (SAJO), Szombathely (SZHE) at an interval of 0.02 seconds (see Fig.~\ref{fig:freq_meas} for locations). As shown in Fig.~\ref{fig:freqdist} (a), the bulk behavior of the frequency [denoted by $f(t)$ to distinguish it from $\omega(t)$ in the Kuramoto rotating frame] basically follows a Gaussian distribution, only that the tail parts are slightly heavier than the Gaussian, which could be better fitted with a $q$-Gaussian with $q$ very close to 1. Due to synchronization of the power system, frequency time series from different locations usually seem to be almost identical on a coarse time scale, regardless of the measured locations~\cite{gorjao2022} and hence give rise to an almost identical bulk distribution, which may be utilized as an initial condition for the Kuramoto equation~\eqref{kur2eq}.

\begin{figure*}[!htbp]
    \centering
    \includegraphics[width=0.99\textwidth]{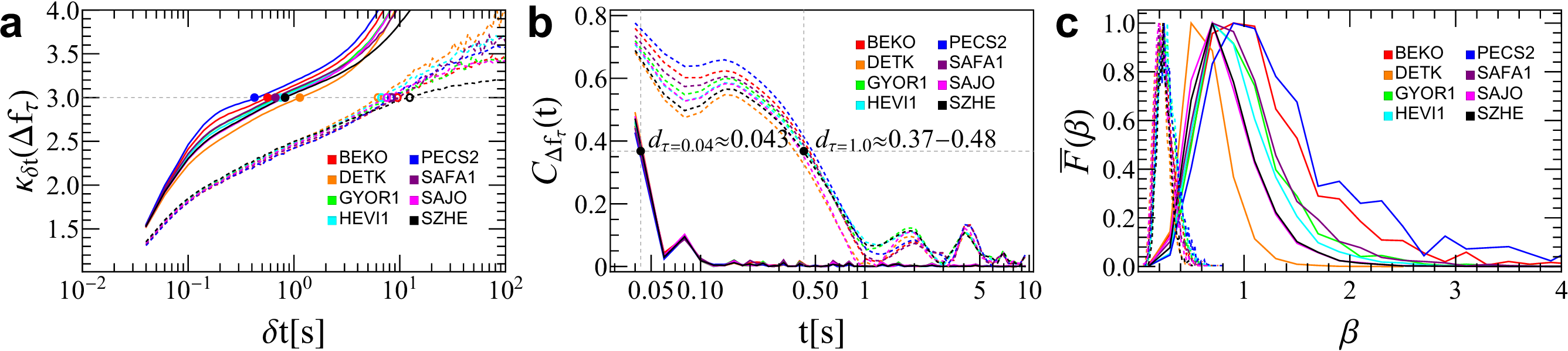}
    \caption{(a) Snippet kurtosis, (b) connected auto-correlation functions, and (c) spectra of frequency increments of different locations at time lags $\tau=0.04$ seconds (solid curves) and $\tau=1$ second (dashed curves). In (a), the super-statistical time scales are $T_{\mathrm{BEKO}}=0.562688$, $T_{\mathrm{DETK}}=1.13538$, $T_{\mathrm{GYOR1}}=0.680575$, $T_{\mathrm{HEVI1}}= 0.748079$, $T_{\mathrm{PECS2}}=0.423406$, $T_{\mathrm{SAFA1}}=0.668517$, $T_{\mathrm{SAJO}}=0.821327$, and $T_{\mathrm{SZHE}}=0.822878$ when $\tau=0.04$, and are $T_{\mathrm{BEKO}}= 9.50441$, $T_{\mathrm{DETK}}=6.21044$, $T_{\mathrm{GYOR1}}=8.35081$, $T_{\mathrm{HEVI1}}=6.68289$, $T_{\mathrm{PECS2}} =8.2801$, $T_{\mathrm{SAFA1}}=8.44334$, $T_{\mathrm{SAJO}}=7.52667$, and $T_{\mathrm{SZHE}}=12.4069$ when $\tau=1$.}
    \label{fig:superstat}
\end{figure*}

However, on a short time scale of a few hundred milliseconds, frequencies fluctuate around the synchronous behavior, leading to distinct dynamics for each recording~\cite{gorjao2021}. These fluctuations can be studied by the increments of frequencies~\cite{tabar2019,gorjao2021}
\begin{equation}
    \Delta f_\tau(t)=f(t+\tau)-f(t)\,,
\end{equation}
where $\tau$ is the incremental time lag. The increment of a time series is useful in that it eliminates the deterministic trends and focuses on the stochastic characteristics of the fluctuations on the shortest time scales. In contrast to the frequency distribution, Fig.~\ref{fig:freqdist} (b) shows that the frequency increment distributions, here with the example for the BEKO case, are characterized by much heavier tails than that of Gaussian distributions (leptokurtic with kurtosis $\kappa>3$). This non-Gaussian behavior hints that the system is genuinely non-equilibrium as power generation and consumption change over a long time scale, e.g.~over one day. Then, similar to the super-statistics of the overall energy distribution of a non-equilibrium system contained in a volume, which can be regarded as a superposition of Gibbs distributions pertaining to equilibrium states reached in small local cells~\cite{beck2003}, it has been demonstrated that the observed non-Gaussian distribution for frequency increment may also be explained by super-statistics if one divides the increment time series into snippets~\cite{gorjao2021}.

Following Ref.~\cite{gorjao2021}, we briefly present the main idea of super-statistics and the results for the Hungarian data-sets. Fig.~\ref{fig:freqdist} (b) showed us that the Kurtosis of the whole frequency increment time series is larger than 3 (i.e.~leptokurtic). Now if one looks at a shorter increment time series of time length $\delta t$, it should be expected that at a short enough intermediate time scale $\delta t=T$, local equilibrium characterized by a Gaussian distribution will be reached. For even shorter increment time series with $\delta t<T$, larger deviations from the mean do not have sufficient chances to occur, so the distribution is intrinsically platykurtic, characterized by thinner tails ($\kappa<3$). In practice, one may slice the whole increment time series corresponding to the time lag $\tau$ into snippets of time duration $\delta t$ and study their averaged kurtosis at this time scale $\delta t$:
\begin{equation}
    \kappa_{\delta t}(\Delta f_{\tau})=\left\langle \frac{\frac{1}{\delta t}\sum_{i=(j-1)\delta t+1}^{j\delta t}\Delta f_{\tau}^4(i)}{(\frac{1}{\delta t}\sum_{i=(j-1)\delta t+1}^{j\delta t}\Delta f_{\tau}^2(i))^2}\right\rangle_{\delta t}\,.
\end{equation}
At an intermediate time scale $\delta t=T$ when $\kappa_{\delta t=T}=3$, the snippets just resemble the local equilibrium cells of non-equilibrium thermal systems, only that the underlying equilibrium distribution is now assumed to be Gaussian, and in accord with super-statistics~\cite{beck2003,gorjao2021}, the frequency increment distribution can be expressed as superposed by a spectrum of Gaussian distributions
\begin{equation}
    p(\Delta f_\tau) = \int_ 0^{\infty} F (\beta) p_N\left(\Delta f_\tau | \beta \right) d\beta\,,
\end{equation}
where $p_N (\Delta f_\tau | \beta) = \sqrt {\frac{\beta}{2\pi}} e^{-\frac {1}{2}\beta \Delta f_\tau^2}$. To properly separate different time scales, a caveat to note is that this super-statistics analysis is valid only if the super-statistical variation time scale $T$ is much greater than the local relaxation time scale $d$, as strong auto-correlation will prevent local equilibrium from being reached. The time scale $d$ can be determined by the auto-correlation function
\begin{equation}
    C(t-t')=\left\langle (\Delta f_{\tau}(t)-\langle \Delta f_{\tau} \rangle)(\Delta f_{\tau}(t')-\langle \Delta f_{\tau} \rangle)\right\rangle
\end{equation}
via $C(d) =e^{-1}C(0)$.

In Figs.~\ref{fig:superstat} (a) and (b), we see that the kurtosis indeed shows a transition from $\kappa_{\delta t}<3$ to $\kappa_{\delta t}>3$ as $\delta t$ increases, and for $\tau=0.04$, $T\gg d$ is strictly held. Fig.~\ref{fig:superstat} (c) further shows the spectrum $\overline{F}(\beta)=F(\beta)/\max(F(\beta))$ with $\beta$ values extracted from the inverse variance of the snippets at $\delta t=T$: 
\begin{equation}
    \beta_T(t)=\frac{1}{\langle \Delta f_\tau^2(t)\rangle_T-\langle \Delta f_\tau(t)\rangle_T^2}\,.
\end{equation}
For $\tau=0.04$ seconds, since $T\gg d$ holds, the spectra $\overline{F}(\beta)$-s span wide distributions and are distinct for recordings of different locations. When we increase the time lag to $\tau=1$ second, $T$-s become only a few times larger than the corresponding $d$-s, and the spectra are much narrower.

What is more, as discussed in Ref.~\cite{gorjao2022}, the frequency increment time series could also show spatial-temporal correlations. Hence, adequate simulations should extend their analysis far beyond uncorrelated noise to correlated noise. Moreover, as demonstrated above for the Hungarian empirical data, it would be interesting to examine if fluctuations in simulations can also lead to super-statistical behavior.

\subsection{Heterogeneity in capacity distributions}

Representing the heterogeneous nature of the transmission infrastructure is typically done either by using the grid model of a country of synchronous area or by modifying the coupling strengths of the Kuramoto equation according to a pre-specified probabilistic distribution. In this section, a short overview is given on both aspects and some shortcomings of the literature are addressed.

Olmi et al. used the representations of the Italian \cite{olmi2014hysteretic} and the German \cite{taher2019enhancing} power grids, which were good examples of heterogeneous topologies; however, the coupling strength was assumed to be identical for all lines. Similarly, a uniform coupling strength, \SI{1600}{MW}, representing the capacity of a \SI{380}{kV} line was used in the examinations of Menck et al. \cite{menck2014dead}. Sch{\"a}fer et al. \cite{schafer2019dynamical} generated two homogeneous and a heterogeneous representation of the Turkish power grid, using the magnitude of power flows to represent the coupling strengths.

Real-world, heterogeneous grid topologies were used by Nishikawa and Motter \cite{nishikawa2015comparative}, who showed that there is a non-trivial structure in the coupling among the generator nodes and that the coupling strength spans across many orders of magnitude. The authors of the present paper generated large synthetic networks with characteristics of real power grids exhibiting hierarchical modular structure, low clustering and topological dimensions, which resemble medium- and low-voltage distribution networks, to examine synchronization processes \cite{PhysRevE.98.022305}.

Kim et al. showed a heterogeneous distribution of load and generation in their paper \cite{kim2019structural}, which were investigated under the condition of varying coupling strength, concluding that concentration of power generation at a single location is likely to increase vulnerability to perturbations. As a case study of the German power grid, they also reveal that the modular structure of the power grid does affect its vulnerability \cite{kim2021modular}. Findings of Ódor and Hartmann went even further, when they presented synchronization phenomena on power grid representations that show heterogeneity not only in the coupling strength but also nodal behavior \cite{odor2020}. They conclude that too weak quenched heterogeneity is not sufficient for power-law distributed cascades, but too strong heterogeneity destroys the synchronization of the system.

\begin{figure}[!htbp]
    \centering
    \includegraphics[width=1\columnwidth]{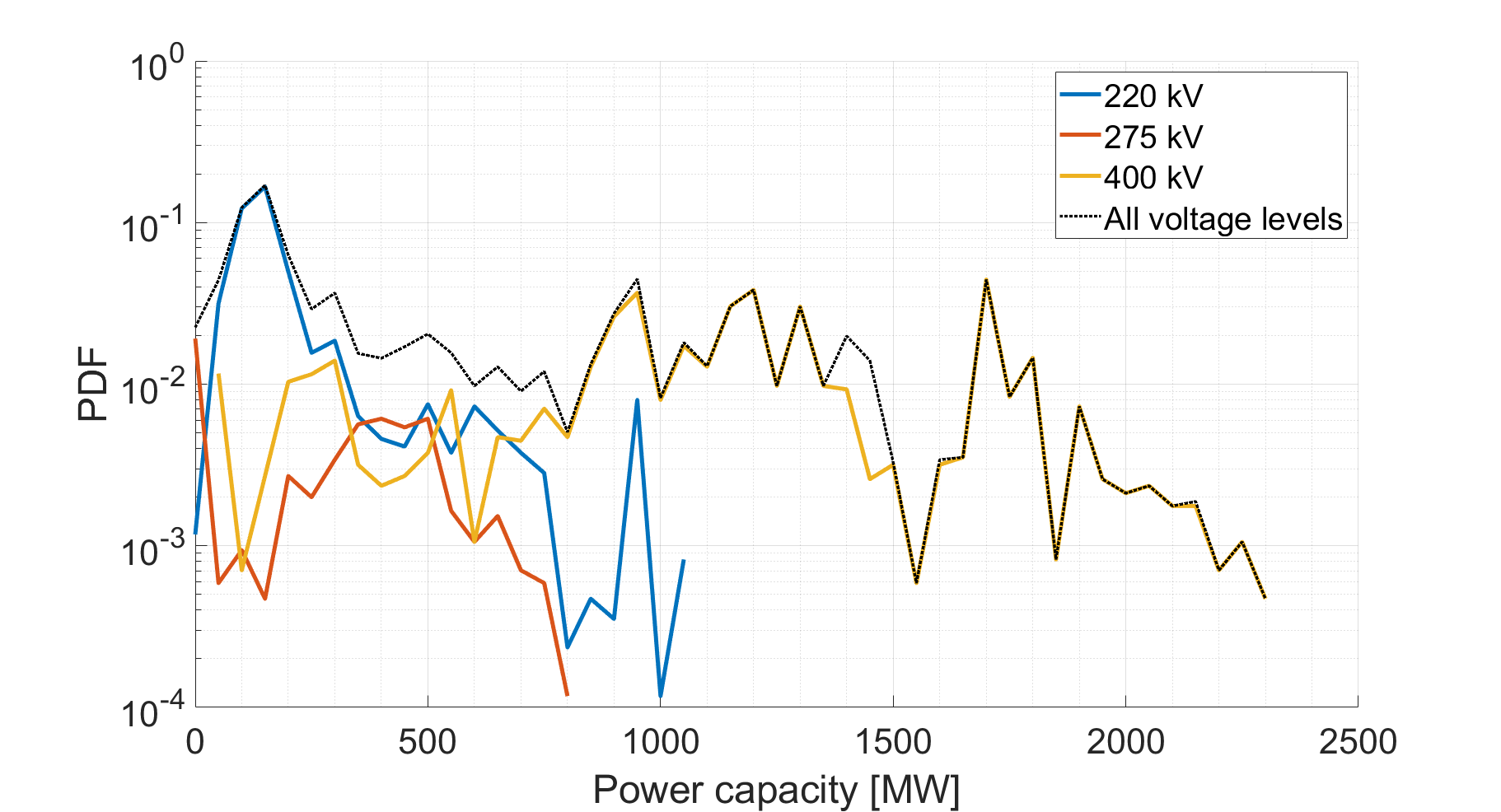}
    \caption{Distribution of thermal power limits of transmission lines included in the ENTSO-E 2016 database. Contributions of 220, 275 and \SI{400}{kV} lines are highlighted as the biggest populations. The distribution does not reflect the combined length of voltage levels, only the number of lines.}
    \label{fig:line_cap}
\end{figure}

In models, motivated by brain function, quenched network heterogeneity has also been shown to cause dynamical criticality, i.e. Griffiths Phases~\cite{Griffiths}, in extended control parameter regions \cite{Munoz2010,Odor2012,MM,HMNcikk,Frus,Cota2016,GProbcikk,Li2017,CCrev,Flycikk}. Also, approaching from the field of network science, very recently, Sánchez-Puig et al.~used random Boolean networks to show that heterogeneity (in time, structure and function) extends the parameter region, where criticality is found \cite{e25020254,sanchez2023heterogeneity}.
To demonstrate the nature of heterogeneity in terms of power capacity, we analyzed the ENTSO-E 2016 data-set, which includes information about the voltage level and the thermal power limit of the transmission lines. Using these two, the theoretical maximum transmittable power can be determined. The above-mentioned data was available for 8511 lines, of which \num{4024} were \SI{220}{kV} lines, 592 were \SI{275}{kV} lines and \num{3738} were \SI{400}{kV} lines; these three groups represent the vast majority of infrastructural elements to be considered. As it is shown in Fig.~\ref{fig:line_cap}, the capacity distribution of three selected voltage levels display rather different characteristics, with mean capacities at 254, 480 and 1144 MW for 220, 275 and \SI{400}{kV}, respectively. The average power capacity of the lines in the ENTSO-E 2016 model, considering all voltage levels is \SI{666.4}{MW}---an atypical value for any widely used voltage level. This also suggests that using a homogeneous capacity in synchronization studies may be misleading as it could easily under- or overestimate the strength of the coupling between nodes.

\begin{figure}[!htbp]
    \centering
    \includegraphics[width=1\columnwidth]{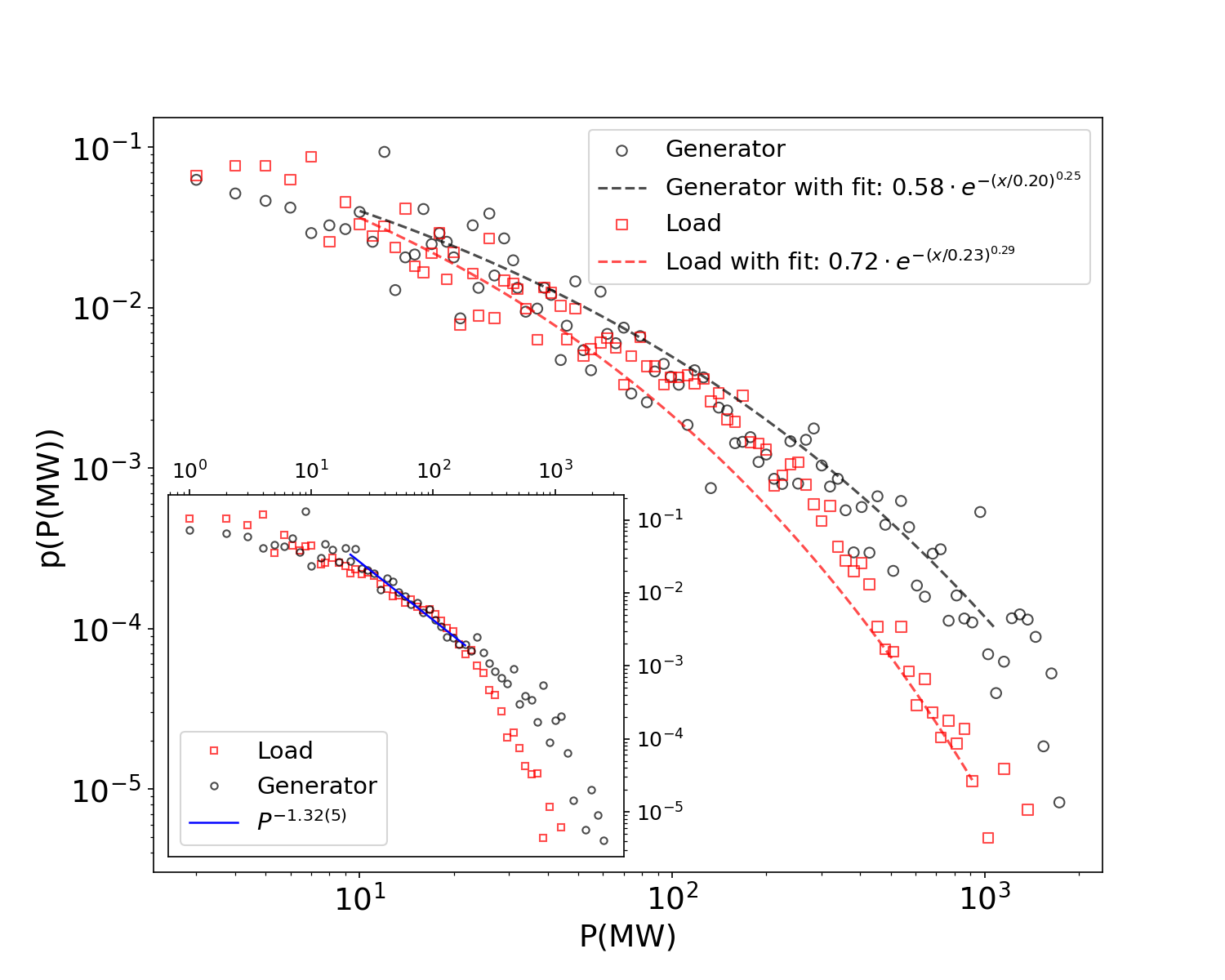}
    \caption{Distribution of nodal generations and loads of the ENTSO-E 2016 database. 
    Power-law fits were applied to the [20...300] MW range in the inset figure. The exponents of the fits are: $y=1.16(5)$ 
    both for generation and load curves, respectively. The load data shows an earlier size 
    cutoff, which is an important characteristic of traditional power systems, where energy is produced 
    in a centralized manner by large power plants to increase efficiency, and energy is consumed in a distributed manner. The main figure shows the same data, with stretched exponential fits, according to 
    Eq.~(\ref{eq:cap-dist}) in the range [10...1000] MW}
    \label{fig:genload}
\end{figure}

\begin{figure}[!htbp]
    \centering
    \includegraphics[width=1\columnwidth]{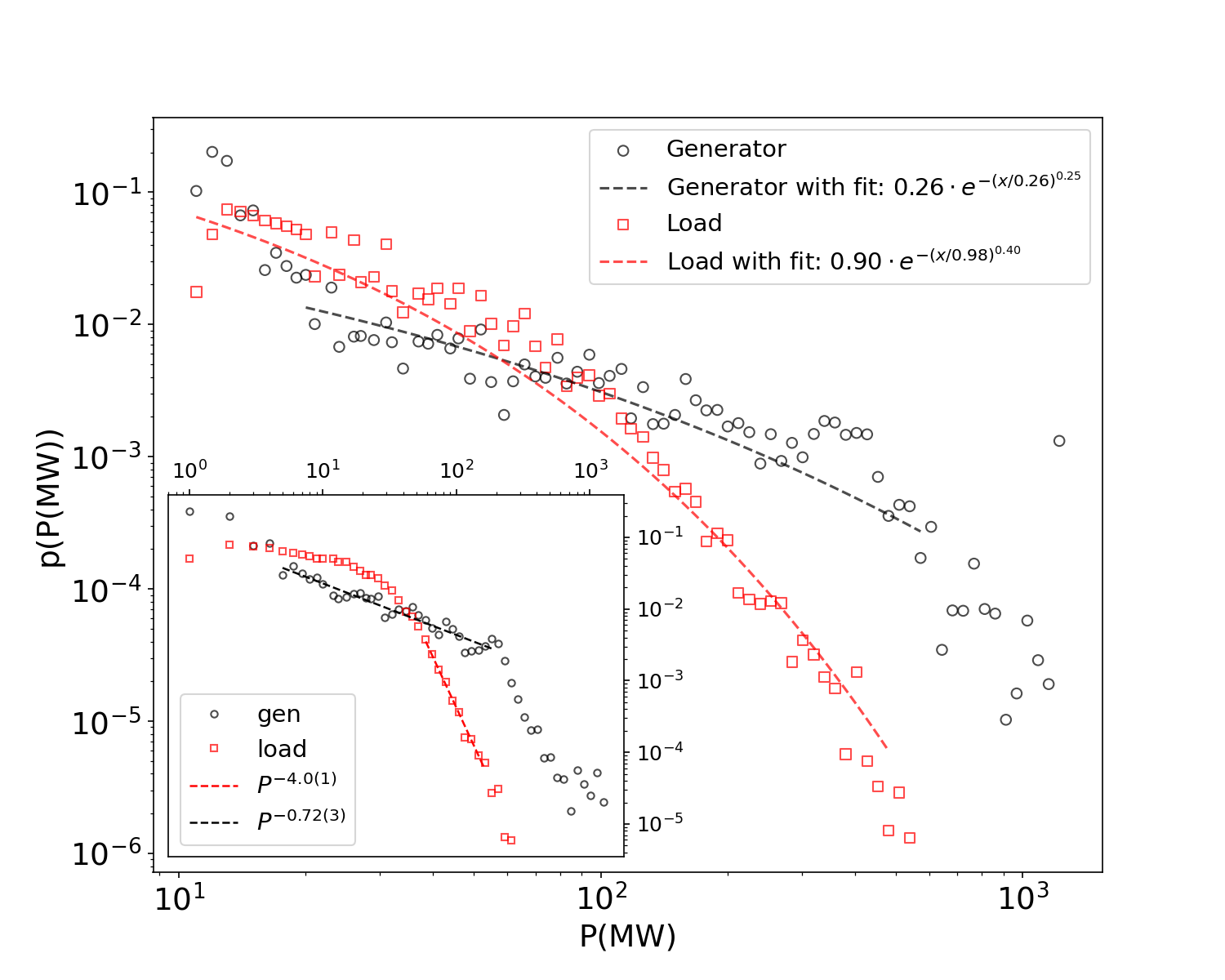}
    \caption{Distribution of nodal generations and loads of the 2021 US~\cite{yixing_xu_2021_4538590} database. 
    Inset: different power-law fits were applied to the [5...200] MW for generators and [50...200] MW for loads.
    The load data shows an earlier size cutoff as for the European case.
    The main figure shows the same data with stretched exponential fit according to Eq.~(\ref{eq:cap-dist}) in the range [20...500] MW. }
    \label{fig:genload-US}
\end{figure}

To demonstrate the nature of heterogeneity and universality in terms of nodal behavior, we present generation and load values of the ENTSO-E 2016 and the 2021 US~\cite{yixing_xu_2021_4538590} data set. 
It has to be noted that these values represent a single, but characteristic operational point of the European and the US power system. As one can see in Figs.~\ref{fig:genload} and \ref{fig:genload-US}, nodal behavior is far from being uniform and can be approximated by stretched exponential functions in wide MW ranges of the form
\begin{equation}\label{eq:cap-dist}
    p(k) \propto \exp(-(k/B)^{\beta}) \ ,
\end{equation}
with similar exponents. But PL-s can also be fitted for rather wide power ranges, with different success and exponents. The generator distributions stretch to higher MW values than the load distributions and exhibit a sharp cutoff if we assume PL-s. We have also tried to fit Pareto like functions without success. These findings support the need for heterogeneous modeling and proper handling of nodal behavior in synchronization studies and universality, which we obtain in other characteristics as well.

\subsection{Topological structure of the European and North American HV power-grids}

\begin{figure}[!ht]
    \centering
    \includegraphics[width=\columnwidth ]{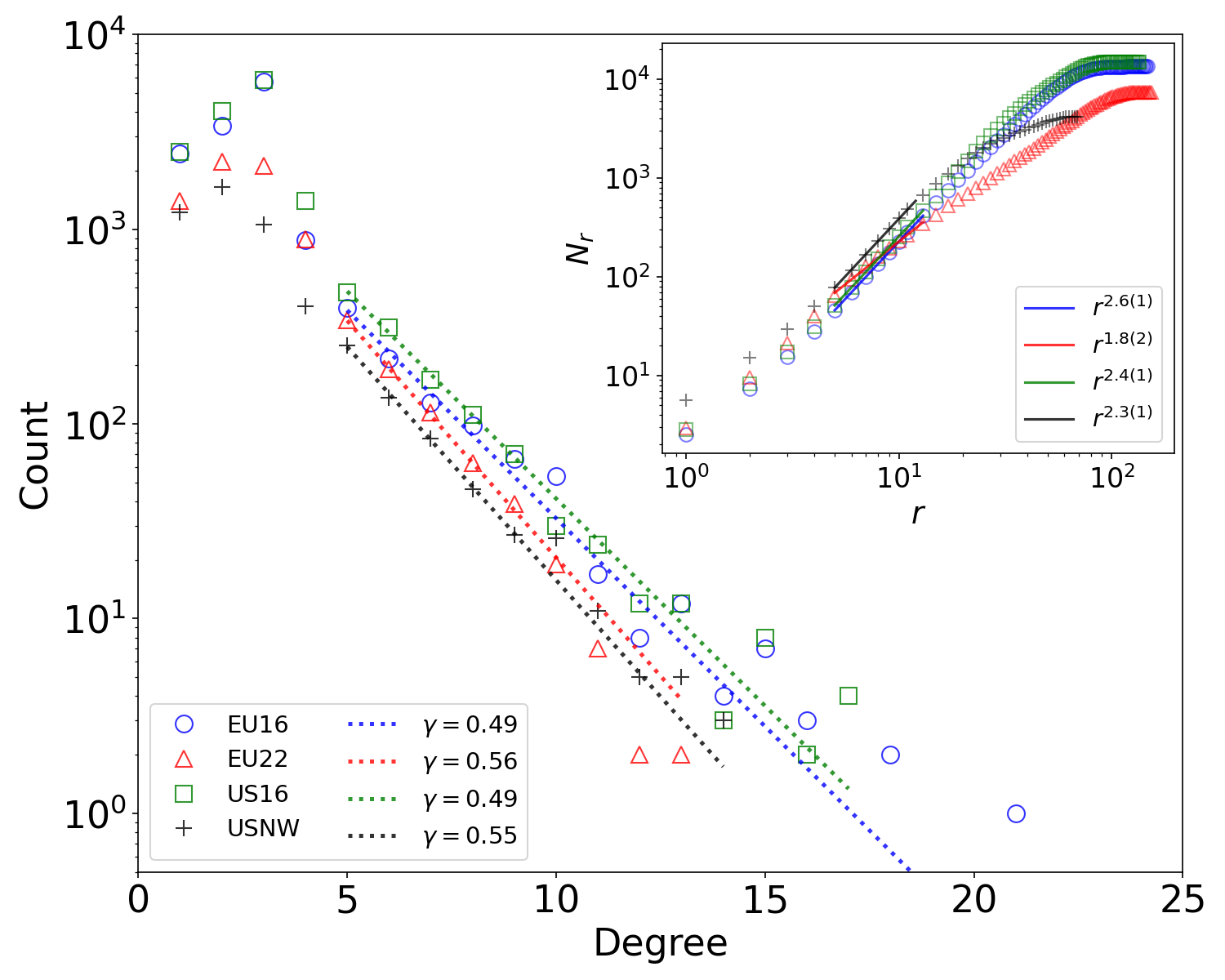}
    \caption{Basic graph invariants of the power grids investigated. 
     Main plot: degree distribution, using logarithmic binning. 
     Exponential fits of form Eq.(\ref{eq:deg-dis}) for $5<k<15$
     resulted in similar $\gamma$ values.
     In the inset we present the graph dimension analysis results of these graphs. PL fits of form Eq.~\ref{eq:Nr} for $5<r<20$ provide similar $d$-s, except for the EU22 case, plotted by red triangles.
    }
    \label{fig:degree}
\end{figure}

As a standard characterization of networks, first we show the comparison of the degree distributions of the
EU16 (European 2016), US16 (North-American 2016), EU22 (European 2022), USNW (US North-West) graphs. 
Fig.~\ref{fig:degree} summarizes the results, obtained using logarithmic binning for the different grids. 
The PDFs look rather similar, with the outlier EU22 case, which decays almost as fast as the
USNW, which is just the subset of the US power-grid: the standard North American HV grid, used in many graph
theoretical papers~\cite{USpg}.
Whether exponential, power-law or mixed distributions are the best fits to cumulative probability distributions of node degrees is still somewhat controversial, as pointed out in \cite{hartmann2021searching}.
Here we present results of the common census for HV networks, fitting with exponential functions
in the form:
\begin{equation}\label{eq:deg-dis}
    p(k) \propto \exp(-\gamma k) \ .
\end{equation}
Note that other authors~\cite{Pgridtop,Albert2002} often consider the cumulative distributions 
$p(k > K)$ to achieve lower fluctuations in the tails.
As we can see on Fig.~\ref{fig:degree}, the $\gamma$ parameters are rather close to the value 
$\gamma=0.5$. They are summarized in Table~\ref{table:spec_quant}.

In what follows, we further probe the topological heterogeneity of power-grid networks through their 
community structures. Communities in networks are usually groups of nodes that are more densely connected to 
each other than to the rest of the network. While several community detection methods exist, they split into 
hierarchical and non-hierarchical methods. Hierarchical methods build a hierarchy of communities by 
recursively dividing the network into smaller and smaller subgroups, while non-hierarchical methods 
directly assign nodes to communities \cite{FORTUNATO201075,Deritei_2014}.
Communities can be considered as unique patterns or the heterogeneity that characterizes the topology.
Naively, one may think the more communities a network has, the more inhomogeneous it is,  
leading to weaker synchronization on the global level.
However, due to the size dependence of $R$ in the case of the crossover synchronization transition 
of the second order Kuramoto model, small communities synchronize at smaller couplings 
$K$~\cite{2D3DKurcikk}, leading to Chimera states, as we show in Sect.\ref{sec:Chim}.

We used openly available data for the power grid network from 
\hyperlink{https://transmission-system-map.entsoe.eu/}{ENTSO EU transmission data set}
from 2016 and from 2022, combined with OpenStreetMap for power-line identifications. 
{\color{black}
For detecting the community structure, we chose the hierarchical Louvain \cite{Blondel2008} method for its speed and 
scalability, this algorithm runs almost in linear time on sparse graphs. Therefore it can be useful on generated test networks with an increased size. The Louvain algorithm is based on modularity optimization. 
For finding communities on a higher level, we also used the Leiden  \cite{TraagWaltmaVanEck2018_LouvainLeiden} algorithm optimizing an extended modularity quotient with a resolution parameter. Which, in this case apparently, or perhaps with highly modular, sparse networks, 
does not give better results than the Louvain method.
The modularity quotient of a network is defined by \cite{Newman2006-bw}
\begin{equation}
Q=\frac{1}{N\av{k}}\sum\limits_{ij}\left(A_{ij}-\Gamma
\frac{k_i k_j}{N\av{k}}\right)\delta(g_i,g_j),
\end{equation}\label{eq:Q}
where $A_{ij}$ is the adjacency matrix, $k_i$, $k_j$ are the degrees of nodes
$i$ and $j$ and $\delta(g_i,g_j)$ is $1$ when $i$ and $j$ were found to 
be in the same community, or $0$ otherwise. $\Gamma$ is the resolution parameter that allows a more generalized community detection, merging together smaller communities. 

Community detection algorithms based on modularity optimization are believed to get 
the closest to the true modular properties of the network. 
With $\Gamma = 1$ we found $\approx$ 425 communities, with maximum modularity score of 
$Q_{EU16} = 0.92724$. For reference we compared the results with the 2016 USA network 
(obtained similarly), it has larger number of nodes in the giant component: \num{14990} 
connected by \num{20880} links. At $\Gamma = 1$ for USA we obtain 460 communities giving high 
modularity score of $Q_{US16} = 0.92525$. This result is in concordance with the 
previously calculated modularity \cite{USAEUPowcikk}.
To obtain community structures similar to the real TSO areas, 
of typically 10-12 domains we have rerun the analysis with $\Gamma < 1$.
Good agreement has been found as discussed in Sect.\ref{sec:geo}  and
shown on Figs.~\ref{fig:comm_size},\ref{fig:16_all},\ref{fig:16_above120},\ref{fig:EU22_10}.

We could also compare the 2016 power-grids with that of a 2022 EU one.
This network, at first glance, seemed to be multiple connected, containing sub-networks 
of different voltage levels. After the unification at nodes with the same node IDs. 
we obtained a graph, which seems to be incomplete in several ways. 
It does not contain nodes with $k=27$ as the 2016 one, but $k_{max}=14$, similarly to 
the USNW, which is just a part of the US system. 
Furthermore, looking at the node degree and edge length distributions, it appears that 
links are missing from the middle $k$ region. More importantly, 
the node number of the largest component is just $N_{EU22}=7.411$, contrary to the 
$N_{EU16}=13.478$, even though the graphical map shows nodes in North Africa as well 
as in the Middle East, see Fig.\ref{fig:EU22_10}.
So, care should be taken about the faithfulness of EU22. 
We just show it for an interesting comparison and to follow the topological changes in
the latest European data.
The graph dimension, measured by the breadth-first search algorithm, defined by
\begin{equation} \label{eq:Nr}
\langle N_r\rangle \sim r^d \ ,
\end{equation}
where $N_r$ is the number of nodes that are at a topological (also called ``chemical'') 
distance $r$ from each other, resulted in $d < 2$, unlike for the other networks,
see Table~\ref{table:spec_quant}. We also show the $N_r$ results in the inset of
Fig.\ref{fig:degree}.

The EU22 is also less heterogeneous than the other networks, calculating the highest modularity 
with $\Gamma = 1$ results in a score of $Q_{EU22} = 0.93346$ from the contribution 
of only 92 communities.
\begin{table}[ht]
\centering
\begin{tabular}{ccccccc}
\toprule
Community &  \makecell{Size \\ (EU22)} & \makecell{$\langle k \rangle$ \\ (EU22)} & \makecell{Size \\ (EU16)} & \makecell{$\langle k \rangle$ \\ (EU16)} & \makecell{Size \\ (US16)} & \makecell{$\langle k \rangle$ \\ (US16)} \\
\midrule
         1 &   924 & 2.72 & 4285 & 2.83 & 3511 & 2.79 \\
         2 &   479 & 2.70 & 2526 & 2.66 & 2829 & 2.98 \\
         3 &  2016 & 2.84 & 1527 & 2.67 & 1640 & 2.72\\
         4 &   698 & 3.06 & 1461 & 2.72 & 1484 & 2.69 \\
         5 &   595 & 2.94 & 1455 & 2.69 & 1396 & 2.93 \\
         6 &  1059 & 2.66 & 966 & 2.77 & 1165 & 2.58 \\
         7 &  1237 & 2.68 & 638 & 2.57 & 768 & 2.97 \\
         8 &    16 & 2.81 & 289 & 2.06 & 710 & 2.57 \\
         9 &   332 & 2.18 & 277 & 2.99 & 673 & 2.70 \\
        10 &    55 & 2.74 & 26 & 3.07 &  390 & 2.84 \\
        11 & - & -&  22 & 3.31 & 230 & 2.43 \\
        12 & - & -&  6 & 2.66 & 194 & 2.69 \\
\bottomrule
\end{tabular}
\caption{Community sizes and average degrees for different data-sets, for the resolution $\Gamma=10^{-4}$.
We refer to sizes here as number of nodes in the respecting community. These structures correspond to the
maps plotted on Figs.\ref{fig:16_all}, \ref{fig:EU22_10}, \ref{fig:USA16}. }\label{tab:comms}
\end{table}
We summarized the community structures of the investigated networks in Table~\ref{tab:comms}.
As we can see, the EU16 and US16 graphs exhibit very similar structures, the same number of communities
for the same resolution $\Gamma=10^{-4}$ and their size distribution is also very similar
as one can see on Fig.~\ref{fig:comm_size}. 
However, the EU22 network is different both in the lower number of communities and their fast decaying
distribution. This strengthens our observation, mentioned above, that the EU22 is an incomplete graph.
We can also see on this graph the effect of $\Gamma$ on the size distributions: lower $\Gamma$ results
in faster decays.

\begin{figure}[!ht]
    \centering
    \includegraphics[width=\columnwidth ]{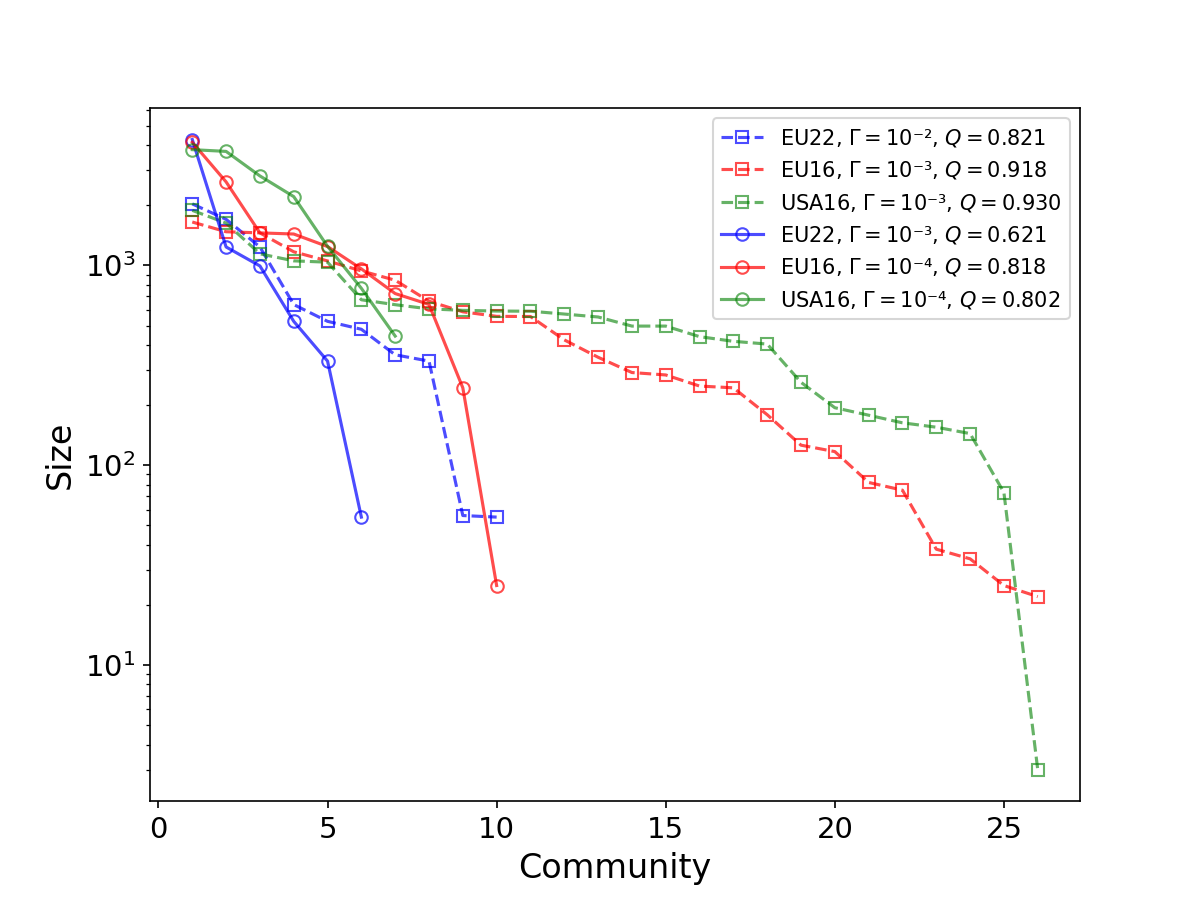}
    \caption{Community size distributions at different $\Gamma$ resolution parameters for
    different networks shown in the legend.
    }
    \label{fig:comm_size}
\end{figure}

\subsection{Topographical structure of the power-grids}\label{sec:geo}

In order to relate topological communities to the topography, we have 
calculated the modularity scores with a lower $\Gamma$ resolution. The results are shown in 
figures Fig. \ref{fig:16_all}--\ref{fig:USA16}.
Note that the whole 2016 data-set (giant component Fig.~\ref{fig:16_all}) and the 2016 one take into account only 
above \SI{120}{kV} lines (Fig. \ref{fig:16_above120}). At the same $\Gamma$, 
shows visibly different community structure, even though only around 50 nodes were removed from \num{13478} 
nodes and the number of links were reduced to \num{17749} not significantly far from \num{18393}. 
The modularity score also shows little difference, however, on the two maps it is noticeable that 
the Apennine Peninsula is split into multiple smaller communities.
It is also worth mentioning that when a higher threshold is introduced (e.g. lines below 
220 kV are left out) the network falls completely apart, considering "true" communities 
with $\Gamma=1$, the resulting modularity score cannot get higher than $Q_{EU16-shattered} = 0.2499$ with 344 communities.
In the case of the 2016 base network with removed links, we also used the giant connected component only.
We compared the community boundaries to the real topology of the European power grid to identify the main topological reasons for the results.

In Fig.\ref{fig:16_all} two cut-sets can easily be seen, the separation of the British Isles and the Iberian Peninsula from the rest of continental Europe. In the first case, the cut-set consists of three HV direct current lines, while in the second case, it is four AC lines (\SI{220} and \SI{400}{kV}). The split in Northern Italy is along the \SI{400}{kV} connection between La Spezia - Vignole - Baggio, an important north-south interconnection. The community in Southern Norway consist mainly of \SI{330}{kV} lines, in contrast to the surrounding areas' \SI{400}{kV} subsystems. Finally, the imprints of history (e.g. the Iron Curtain) can be recognized in Central Eastern Europe, which is separated from the Western part of the continental system, but is also distinguishable from the IPS/UPS (Integrated Power System, Unified Power System) cooperation of the former Soviet Union.

Fig.\ref{fig:16_above120} shows similar results for the British Isles, the Iberian Peninsula and the separation between Central Eastern and Western Europe. A notable difference is that the cut-set between Italy and its neighboring countries is formed according to the political borders in this case. Northern and Southern parts of Italy are separated along bottlenecks formed by the Piombino - Poggio, Piombino - Calenzano and the Candia - Teramo \SI{400}{kV} lines. The split in Denmark is along the parallel \SI{400}{kV} connections between Ferslev and Jardelund. Finally, the former Soviet states are separated from the rest of Europe along the different transmission voltage levels, which is \SI{330}{kV} in the former and \SI{220} and \SI{400}{kV} in the latter region.

A partially different result of community detection is seen in Fig.\ref{fig:EU22_10}, where the effects of HV direct current lines is more emphasized. This is clearly seen (i) in the case of Ireland and Great Britain with the Auchencrosh - Ballycronanmore (the Moyle Interconnector) and Deeside - Woodland (the East–West Interconnector) lines, (ii) for the island of Sardinia with the Bonifacio - S.Teresa (SACOI) and Fiume Santo - Latina (SAPEI) lines, and (iii) in the Southern Scandinavian region with the Fraugde - Herslev (Storebælt HVDC), Bjæverskov - Bentwisch (KONTEK), Kruseberg - Herrenwyk (Baltic Cable) connections. Other easily identifiable cut-sets are seen on the western borders of Turkey (Filippi - N.Santa single and Maritsa Iztok 3 - Hamitabat double \SI{400}{kV} lines) and between Europe and Africa (Fardioua - Tarifa line).

As for the power grid of the USA (Fig.\ref{fig:USA16}) similar telltale signs are seen. In general, the outlines of the Eastern, the Western and the Texas Interconnection can be recognized, but both large areas are divided into more sub-parts. The northern and southern parts of the Western Interconnection are divided along the cut-set of \SI{500}{kV} lines, with the state of Colorado, lacking significant interconnections, forming a separate community. The three largest communities of the Eastern Interconnection resemble the use of different dominant transmission voltage levels, namely \SI{345}{kV} in the northwestern region, and a mix of \SI{161}, \SI{230} and \SI{500}{kV} in the remaining parts. The other three large ones (two in the US Northeast and Florida) are more the results of their geographical properties. An interesting feature of the map is that the orange community includes both Eastern and Western parts; the backbone of this community is a \SI{230}{kV} topological formation.

In conclusion, while the detection of communities helps in understanding the unique patterns and the heterogeneity that characterizes the topology, knowledge on physics and the physical properties of the underlying grid, expertise in the domain of power system planning and operation can largely contribute to the narrative of the analysis.

\begin{figure}[!ht]
    \centering
    \includegraphics[width=\columnwidth ]{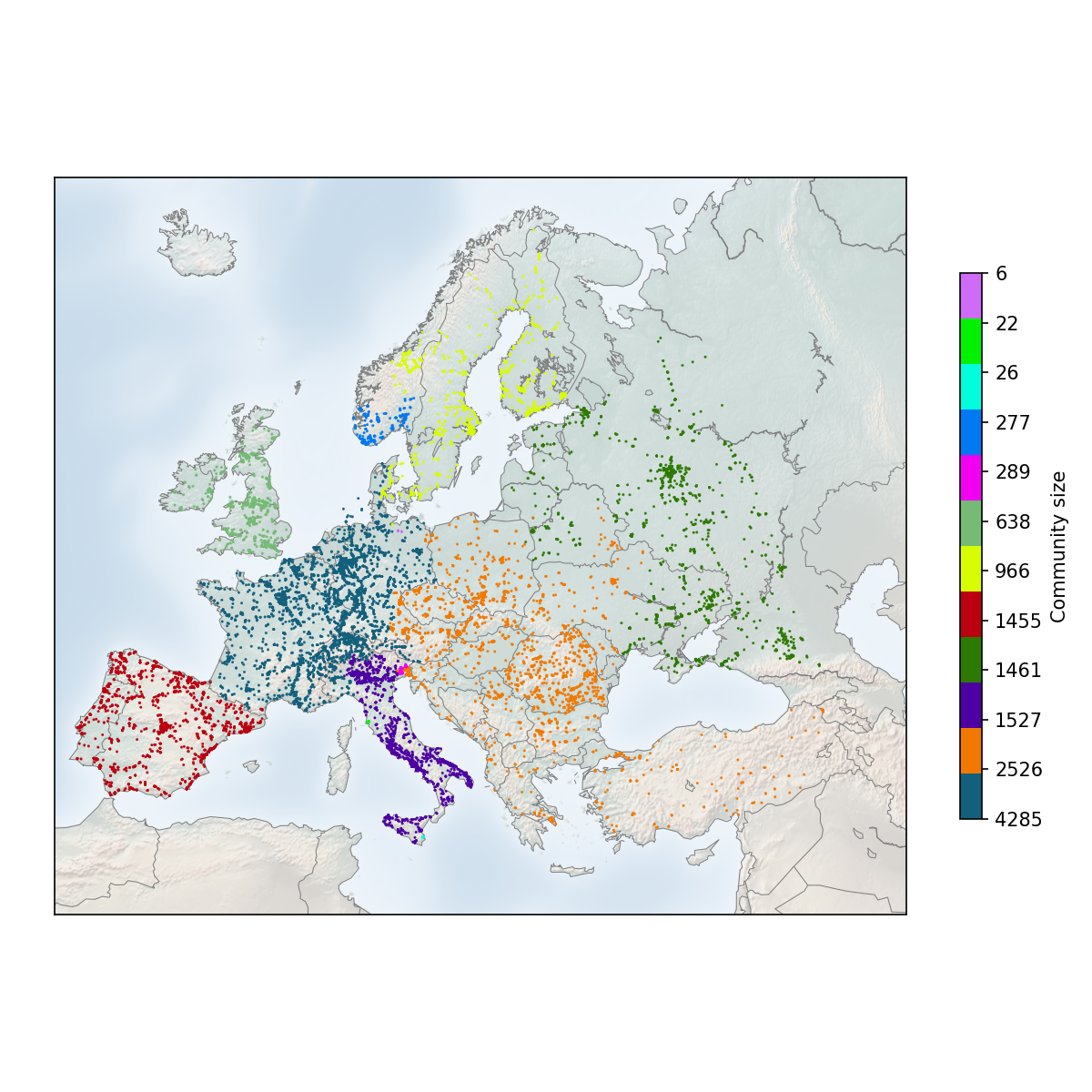}
    \caption{{\color{black}All nodes of the European power-grid 2016 data separated into 12 communities,
    taking into account admittance, using  a giant component of \num{13478} nodes connected by \num{18393} links,
    maintaining the modularity score close to the maximum  $Q \approx 0.795$. }}
    \label{fig:16_all}
\end{figure}

\begin{figure}[!ht]
    \centering
    \includegraphics[width=\columnwidth ]{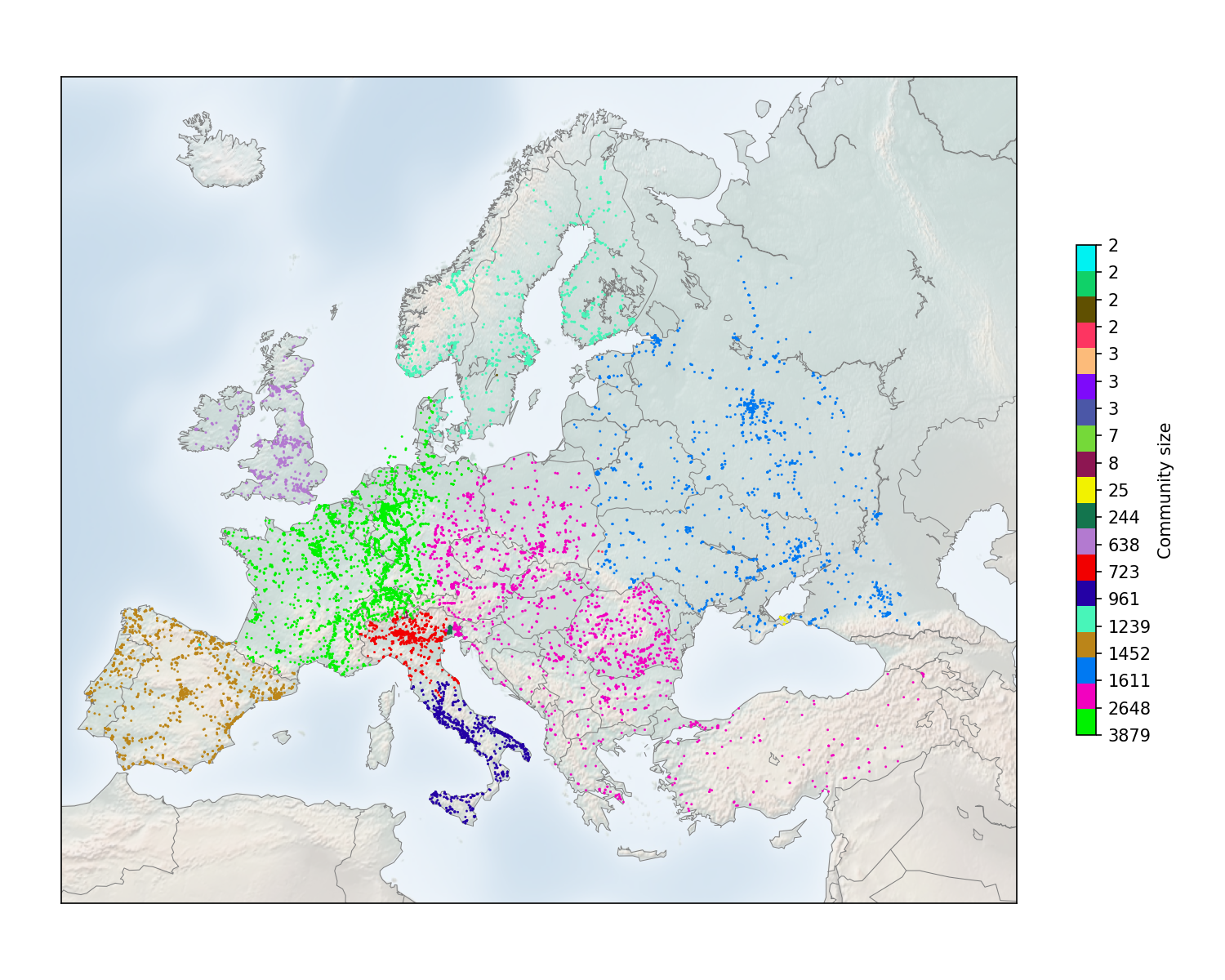}
    \caption{Nodes of the European HV power-grid 2016 data giant component.
    Excluding lines below 120 kV and using $\Gamma=10^{-4}$ resolution,
    taking into account the admittance values the graph is separated into 19 communities.
    A giant component of \num{13420} nodes linked with \num{17749} edges emerges, giving a 
    modularity score at this resolution $Q \approx 0.785$
    }
    \label{fig:16_above120}
\end{figure}

\begin{figure}[!ht]
    \centering
    \includegraphics[width=\columnwidth ]{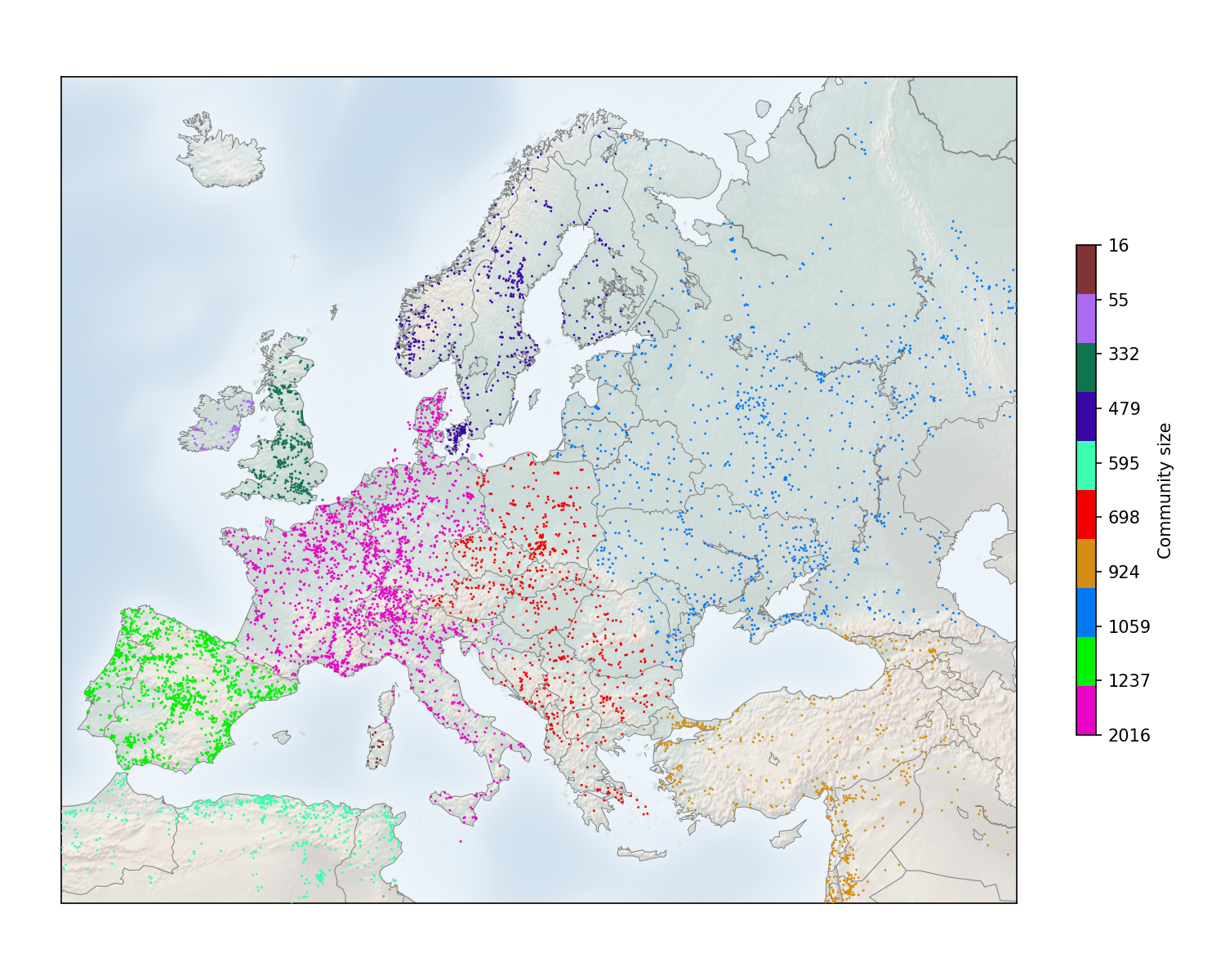}
    \caption{{\color{black}All nodes of the European power-grid 2022 data giant component, separated into 10 communities,
    taking into account the admittances and \num{7411} nodes connected by \num{10912} edges without smaller voltage level 
    edges, maintaining the modularity score  $Q \approx 0.854$. }}
    \label{fig:EU22_10}
\end{figure}

\begin{figure}[!ht]
    \centering
    \includegraphics[width=\columnwidth ]{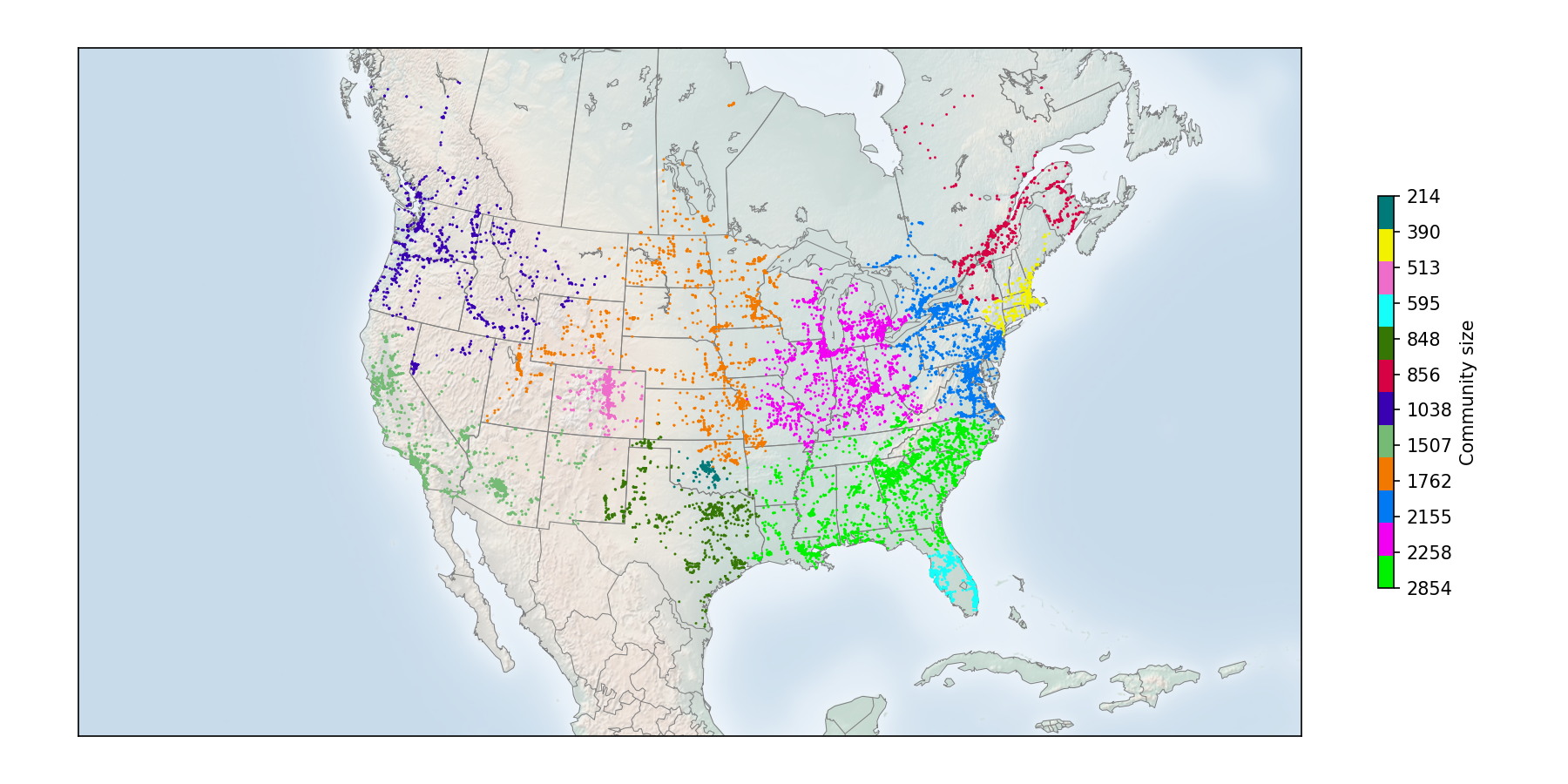}
    \caption{{\color{black}All nodes of the USA power-grid 2016 data giant component, separated into 12 communities,
    taking into account the admittances and \num{14990} nodes connected by \num{20880} edges, maintaining the modularity score $Q \approx 0.859$ with resolution $\Gamma=\num{1e-4}$. }}
    \label{fig:USA16}
\end{figure}

\section{Numerical modeling\label{sec:3}}

To solve the differential equations, in general we used the adaptive
Bulirsch--Stoer stepper~\cite{boostOdeInt}, which provides more precise
results for large $K$ coupling values than the fourth-order Runge--Kutta method.
The solutions depend on the $\omega_i^0$ values and become chaotic,
especially at the synchronization transition, and thus to obtain
reasonable statistics, we needed strong computing resources, using parallel codes
running on GPU clusters. 

To obtain larger synchronization, the initial state is set to be
phase synchronized: $\theta_i(0)=0$, but due to the hysteresis
one can also investigate other uniform random distributions like:
$\theta_i(0) \in (0,2\pi)$.
The initial frequencies were set as: $\dot{\theta_i}(0)=\omega_i^0$.

To characterize the phase transition properties, both the phase order parameter $R(t)$ and the frequency spread $\Omega(t)$, termed the frequency order parameter, is studied. We measured the Kuramoto phase order parameter:
\begin{equation}\label{ordp}
z(t_k) = r(t_k) \exp\left[i \theta(t_k)\right] = 1 / N \sum_j \exp\left[i \theta_j(t_k)\right] \ .
\end{equation}
Sample averages for the phases
\begin{equation}\label{KOP}
R(t_k) = \langle r(t_k)\rangle
\end{equation}
and for the variance of the frequencies
\begin{equation}\label{FOP}
        \Omega(t_k,N) =  \langle \frac{1}{N} \sum_{j=1}^N (\overline\omega(t_k)-\omega_jt_k))^2  \rangle
\end{equation}
were determined, where $\overline\omega(t_k)$ denotes the mean frequency within each respective sample.
To locate the synchronization crossover points better, we have also determined their variances: $\sigma(R)$ and $\sigma(\Omega)$.

\subsection{Modeling admittances and weights in incomplete databases\label{sec:3a}}

In order to carry out synchronization calculations for detailed case studies, missing admittances and graph edge weights have to be estimated. One of the possible solutions is to use the physical parameters of the grid, if they are known. For the following example, the backbone of the used network data is from the \href{https://www.power.scigrid.de/pages/downloads.html}{SciGRID project}, which relies on the 2016 statistics of ENTSO-E and data obtained from OpenStreetMap (.osm) files. Since acquiring data from .osm files is not always possible, the resulting data set may be incomplete. To resolve the problem, assumptions are necessary on how to substitute the missing data.

We can understand the power grid as a graph, where the nodes correspond to generators (power sources) or loads (consumers), while the transmission lines can be considered the edges in the graph. First, we considered the largest connected component of the grid. Selecting the known voltage data for the links makes it possible to estimate the rest. Averaging the available voltages in the giant component yields the average voltage, $\overline V$ of the known links. As the simplest possible assumption, we substituted this value as the voltage level for every unknown entry in the database. 

This is a different approach compared to what is used typically in synthetic grid models; e.g. the original SciGRID database extracts topological data from OpenStreetMap files using an SQL script, while we are building a model dominantly relying on physical properties. Based on our expertise in grid modeling, a proposal can be made on the specific values of the relevant quantities (such as the resistance, the reactance and the capacitance) as the function of the voltage level. For the used specific values see Table ~\ref{table:spec_quant}.
\begin{table}[!ht]
\caption{Characteristic values of relevant physical quantities in the modeled European power grids.} 
\centering 
\begin{tabular}{c c c c c} 
\hline\hline 
Voltage [$\si{\kilo\volt}$] & R\textsubscript{c} [$\si{\ohm/\km}$] & X\textsubscript{c} [$\si{\ohm/\km}$] & C\textsubscript{c} [$\si{\nano\farad/\km}$] & P\textsubscript{c} [$\si{\mega\watt}$]\\ [0.5ex] 
\hline 
120 & 0.0293 & 0.1964 & 9.4 & 170\\ 
220 & 0.0293 & 0.2085 & 9.0 & 360\\
380/400 & 0.0286 & 0.3384 & 10.8 & 1300\\[1ex] 
\hline 
\end{tabular}
\label{table:spec_quant} 
\end{table} 

Electrical parameters could be calculated by grouping the edges in different voltage levels and performing the calculations on the common voltage level of the network. We identify three groups in the networks as shown in Table~\ref{table:spec_quant}: $\SI{120}{\kilo\volt}$, $\SI{220}{\kilo\volt}$, $\SI{380}{\kilo\volt}$. The boundaries between these categories are defined as the arithmetic mean of two neighboring voltage levels and a link will have the characteristic parameters of the category it is closest to. That is, every link below $\SI{120}{\kilo\volt}$ is part of the first category, having the parameters of the $\SI{120}{\kilo\volt}$ lines, every line above $\SI{380}{\kilo\volt}$ is part of the $\SI{380}{\kilo\volt}$ class, and the category of rest of the links will be decided based on the closeness to the above-defined boundaries. Here we have to refer to Section II.B., where it is shown that the voltage levels with the biggest population ($\SI{220}{\kilo\volt}$ and $\SI{400}{\kilo\volt}$) represent approx. 90\% of the lines.

Having decided the voltage class of a line we can compute the relevant quantities as:
\begin{gather}
    R_{ij} = \left(\frac{U_c}{U_{ij}}\right)^2\cdot L_{ij}\cdot R_{c_k}\\
    X_{ij} = \left(\frac{U_c}{U_{ij}}\right)^2\cdot L_{ij}\cdot X_{c_k}\\
    P_{ij} = P_{c_k},
\end{gather}
where $R_{c_k}$, $X_{c_k}$ are characteristic values belonging to level $k$ and $U_{c}=\SI{220}{\kilo\volt}$ is the most common level for the European grid. We considered the weight of the link from node $i$ to $j$ to be defined as:
\begin{equation}
    W_{ij} = \frac{P_{ij}}{X_{ij}}\Big/\left\langle\frac{P}{X}\right\rangle,
\end{equation}
where $P_{ij}$ is the nominal power of the link on its voltage level, and $X_{ij}$ is its impedance. The normalization factor has been chosen to be the average value of this fraction calculated for the whole network.

We have calculated PDFs of admittances of the European EU16, and for comparison, of the US16 
North-American power-grid~\cite{SciGRIDv0.2, wiegmans_2016_47317}, obtained by the same data completion
method as described above. 
By this extension, in the case of the 2016 North-American network 9527 (45\%)
 whereas in the case of the 2016 European and 2022 European networks 5167 (28\%), and respectively 40 
 (0.3\%) new links' $Y$ have been estimated.
As one can see in Fig.~\ref{fig:elo-EU-USA-padm} the US16 and EU16 exhibit very similar heavy tails, which can be fitted by 
PL-s of the form $p(Y) \propto Y^{-x}$ in the region: $10^2 < Y < 10^4$ [1/ohms], characterized by the exponents: $x_{EU16}=2.03(3)$, $x_{US16}=2.05(5)$. 

Furthermore, a Lomax-II \cite{Lomax2}, related to Pareto distribution in the form:
\begin{equation}\label{eq:pareto}
    p(Y)=\frac{A }{\Lambda}  \left[{1+\frac{Y }{ \Lambda }}\right]^{-(A +1)},\qquad Y\geq 0,
\end{equation}
can describe even the low $Y$ range, with a high goodness factor: $R^2 = 0.987$ for EU16 and $R^2 = 0.994$ 
for US16 as shown in the inset of Fig.~\ref{fig:elo-EU-USA-padm}.
However, in case of the EU22 data we see a different behavior in the
intermediate $Y$ range, which is probably the consequence of the incomplete data-set, we have not tried to apply a numerical tail fit.

These heavy tails of lines with larger admittances are the consequence of very short edge lengths in the databases,
which can take values of a few meters.
\begin{figure}[!ht]
    \centering
    \includegraphics[width=\columnwidth ]{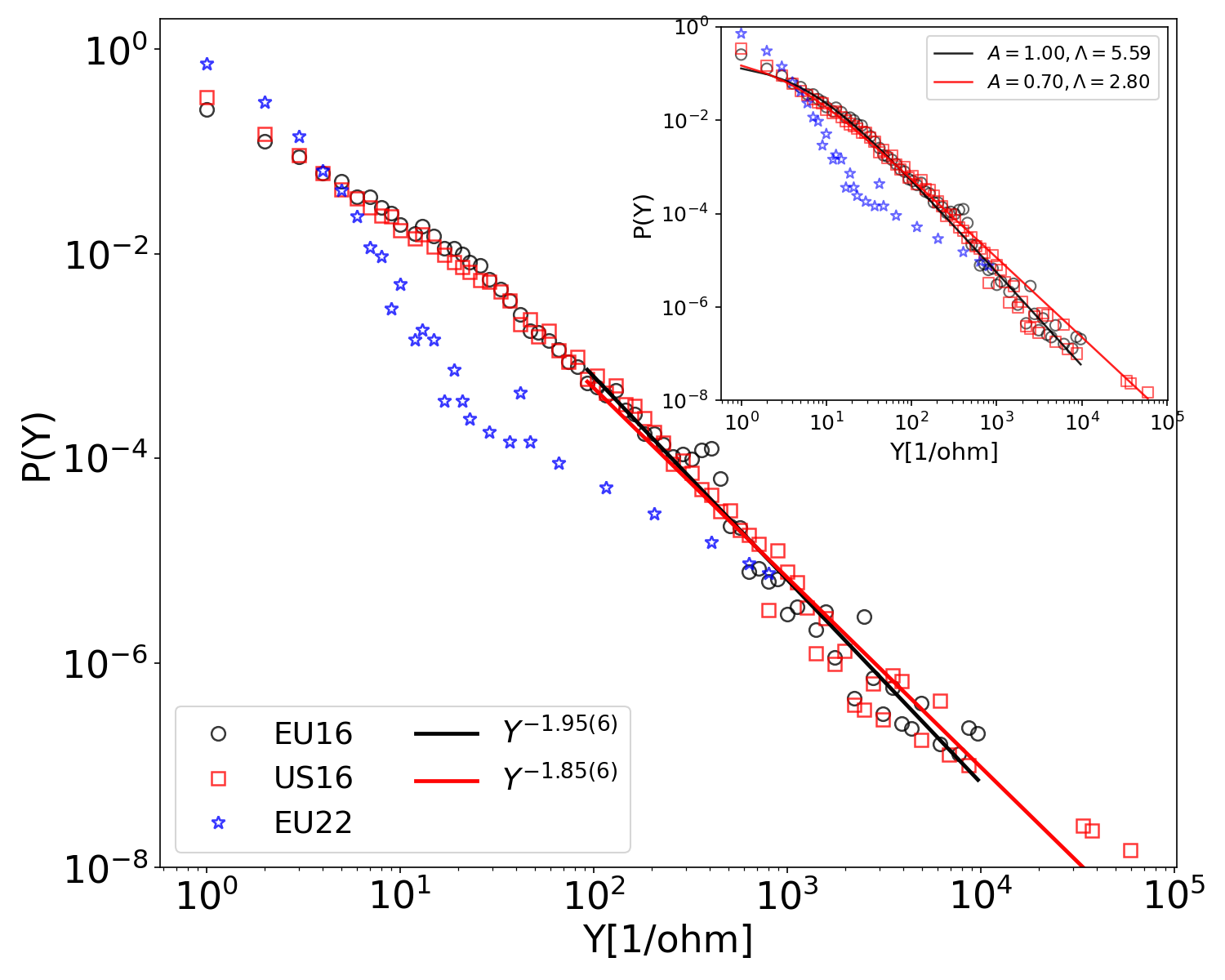}
    \caption{Probability distributions of the calculated admittances of the 2016 European (bullets), North-American (boxes) SciGRID and the 2022 EU (stars) networks. Dashed line: least squares PL fit to the EU16, dotted line: 
    for the US16, in the region: [100..10000] 1/ohm.
    The inset shows the same data, fitted with the form Eq.(~\ref{eq:pareto}), which works well down to the 
    10 [1/ohm] region.}
    \label{fig:elo-EU-USA-padm}
\end{figure}
In case of the EU22 data-set we do not see the same behavior as for the others, probably because it does not 
contain all HV links. 
The incompleteness of the EU22 graph is obvious, because it contains only $10298$ edges as compered to 
the EU16 case, which has $18393$ one, even though the EU22 includes territories of North Africa and the 
Middle East, see Fig.\ref{fig:EU22_10}.

To investigate this further, we have analyzed the cable lengths of the 2016 SciGRID data bases.
Fig.\ref{fig:elo-lengths} shows the PDF-s of $l_\mathrm{max}/l$, which would be proportional
to the admittances, if cables were uniform with the same characteristic resistances.
\begin{figure}[!ht]
    \centering
    \includegraphics[width=\columnwidth ]{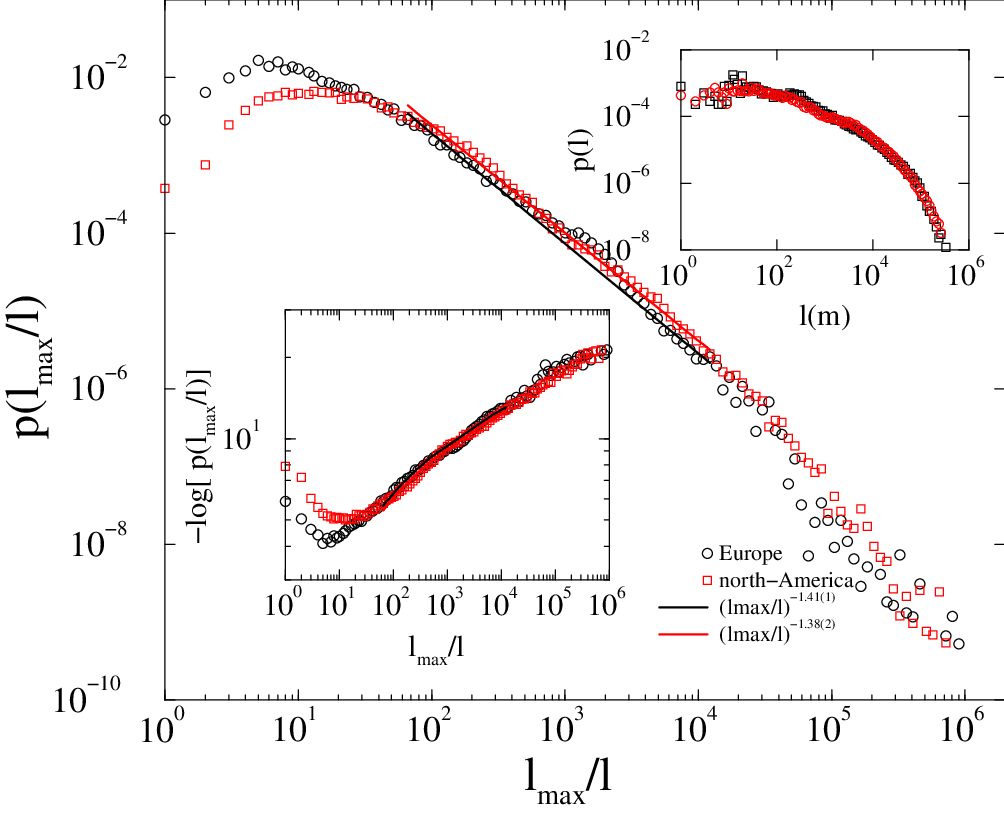}
    \caption{Probability distributions of the inverse of cable lengths of the European and North-American 
    SciGRID networks. Left inset: the same data plotted on the $-\ln(p)$ scale to compare with the stretched 
    exponential assumption, that would correspond to a straight line tail, 
    Right inset: probability distributions of the line lengths in meters. 
    }
    \label{fig:elo-lengths}
\end{figure}
The distributions are very similar, but the tails break down more rapidly 
than for the admittance distributions, suggesting a stretched exponential. The right inset, showing the PDF of lengths in meters, exhibits larger curvatures on the log-log plot.

\subsection{Frustrated synchronization, Chimeras in modules of the European power-grid}
\label{sec:Chim}

Heterogeneity is known to cause so-called frustrated synchronization~\cite{Frus,FrusB},
such that for a given control parameter set certain domains are synchronized, while 
others are not. This is also related to the so-called Chimera states~\cite{chimera,zakharova2021chimera}. 
Now we show that this really happens here by calculating $R(t\to\infty)$ and 
the variance $\sigma(R)$ in the communities determined before.
We used the $i=1,...,12$ community decomposition and determined $R_i(t\to\infty)$
and averaged over the numerical solution of $100$ independent self-frequency
samples and for $1200 < t \leq t_\mathrm{max}=2000$ time steps in the steady state
following an initialization from fully phase synchronized states.
As Fig.\ref{fig:R-K-12} shows, $R_i(t\to\infty)$ grow with $K$ in a different
way in different communities and exhibit distinct fluctuation peaks, 
shown in the inset. That means for a given $K$, some communities are synchronized 
at least partially, while others are not.
Note, that in case of EU16 network, the communities $3,6,7$ the Kuramoto 
order-parameter remains close to $1$ and there is no peak in $\sigma(R)$
\cite{SD-GO-Dresd-proc}.
This is the consequence of containing only a few number of nodes in the communities,
(see Table \ref{table:spec_quant}), which synchronize them for very a small $K$.

On the other hand, in the case of the EU22 graph communities: $1,2$ synchronize at larger $K$-s than the whole system, so the simple size dependence law, which is valid for independent finite graphs~\cite{odor2023} does not seem to hold, but
non-trivial topological features of the EU22 graph cause that the
Turkish and the Nordic region are less synchronized for a given $K$.
This has also been found to be true when we used a larger damping factor: $\alpha=3$ in the solution of the Kuramoto equation.

In the case of disordered initial conditions, the fluctuation peaks shift to the right with respect to the above results and the steady state values are lower than of the global order parameter, which exhibits a hysteresis and meta-stable states near the crossover point~\cite{USAEUPowcikk}.}

We have found similar results for the frequency spread order
parameter $\Omega$, without a peak in the variances $\sigma(\Omega)$
as in previous publications~\cite{USAEUPowcikk,odor2023}.

\begin{figure}[!ht]
    \centering
    \includegraphics[width=\columnwidth ]{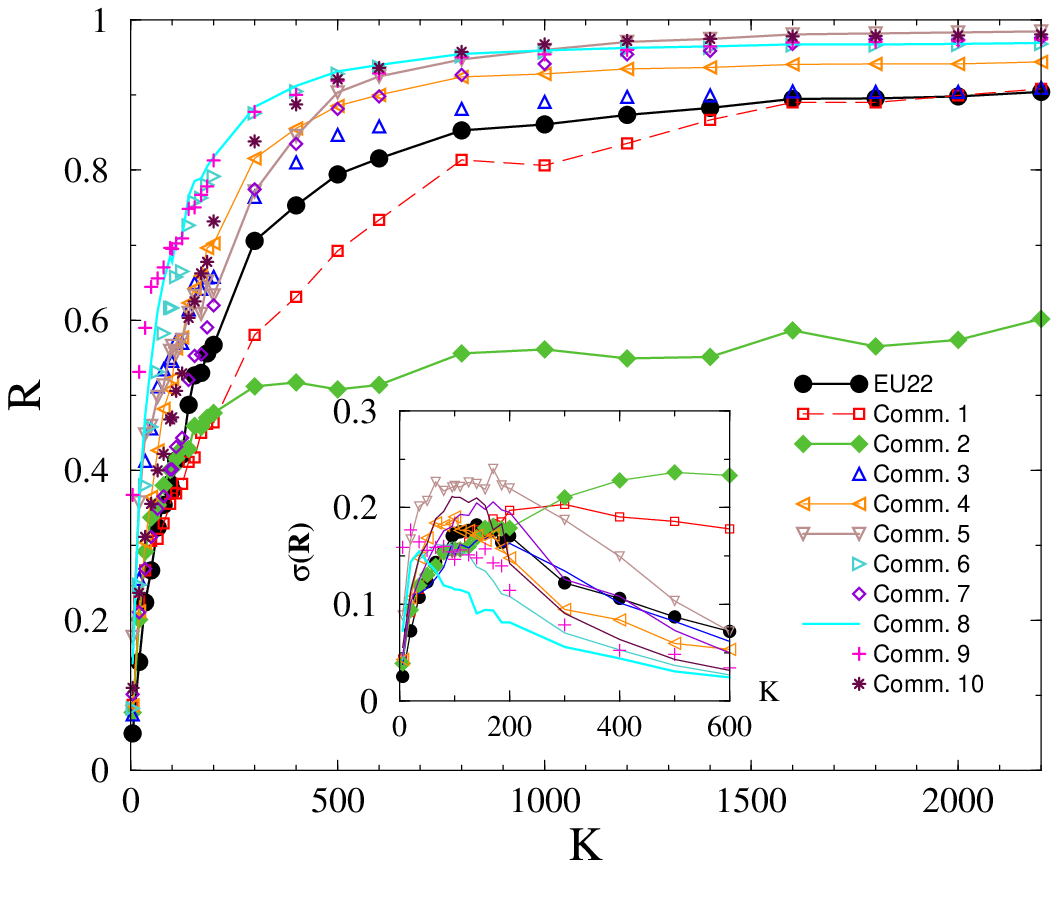}
    \caption{Community dependence of $R$ for different $K$-s at $\alpha=0.4$ in
    the EU22 network shows different phase synchronizations, corresponding to 
    "Chimera" states. 
    The thick black curve denotes the synchronization of the whole system, which grows the slowest by increasing $K$.
    Inset: Fluctuations of the same data, showing different synchronization points.
    The thick black curve, representing the whole system does not exhibit the rightmost peak,
    as for the EU16 power-grid~\cite{SD-GO-Dresd-proc}, instead the community 2, corresponding 
    to the Nordic power-grid.
    }
    \label{fig:R-K-12}
\end{figure}

\section{Discussion \label{sec:5}}

In summary, we have obtained similar graph and electrical characteristics for the North American and European power-grids, i.e. for the graph dimension: $d$, modularity: $Q$,
average degree: $\langle k \rangle$, coefficient of the exponential degree form: $\gamma$, capacitance stretched exponent of the generators: $\beta$, and the PL exponent of cable admittances: $x$ as summarized in Table~\ref{table:sum}. The agreement
is remarkable, although the EU22 values are a bit off, due to the incomplete
data-set, we have discussed in the text.

Besides, we found good agreement in the frequency fluctuation analysis of the Hungarian MAVIR data $q \simeq 1.1$ (entropic index) with the results for the Nordic power-grid see Ref.~\cite{gorjao2021spatio}),

We see an interesting deviation in the community level synchronization behavior of EU22 from the EU16~\cite{SD-GO-Dresd-proc} results and expectations based on synchronization of the second order Kuramoto model on lattices. Namely, the Turkish and Nordic communities synchronize at higher global
couplings than the whole system, in contrast with expectations by the synchronization crossover size dependence of independent, lattice systems~\cite{odor2023}. This is related to the special topological connections of these regions in the EU22 power-grid network and strengthens the necessity to simulate power-grids carefully, by checking approximations in the interactions.

\begin{table*}[!ht]\label{table:sum}
\caption{Summary of fitted data for the power grids. $d$ denotes the graph dimension (\ref{eq:Nr}), 
$Q$ is the modularity quotient (\ref{eq:Q}), $\langle k\rangle$ is the average degree, $\gamma$ is the decay exponent of 
the exponential of the degree distribution (\ref{eq:deg-dis}), $\beta$ is the power capacitance exponent of the
generators (\ref{eq:cap-dist}), $x$ is the admittance decay exponent. } 
\centering 
\begin{tabular}{|c | c | c | c | c | c| c | c | } 
\hline\hline 
Network  & $d$ & Q ($\Gamma=1$) & \makecell{Q($\Gamma=10^{-4}$) \\ 100 averages} & $\langle k \rangle$ & $\gamma$ &  $\beta$ & $x$ \\ [0.5ex] 
\hline 
USNW & 2.3(1) &  0.929  &   -     & 2.67 &  0.55 & - & -\\ 
EU16 & 2.6(1) & 0.927 & 0.829  & 2.729 & 0.49 & 0.25 & 2.03(3) \\ 
US16 & 2.4(1) & 0.925 & 0.734 & 2.785 & 0.49 &  0.25 & 2.05(5) \\
EU22 & 1.8(2) & 0.933 & 0.693 & 2.779 &  0.56 &  - & - \\[1ex] 
\hline 
\end{tabular}
\end{table*}

\section{Conclusions}

We have provided a numerical analysis of the European and North American 
power-grids, both on the network topological level and the dynamical solution of the swing equation description of energy transfer. 
Having the augmented databases, we could assign weights to the network. 
These weights could be fed as additional input to the simulations.  This represents an important enhancement compared to our previous results, where we only considered the network's topology, i.e. we used only the adjacency matrix. \cite{USAEUPowcikk}.

Our non-perturbative
analysis goes beyond linear approximations or DC models and reflect a non-trivial
relation of heterogeneity and synchronization stability. Community structure
decomposition, followed by solution of equations of motion do not only reveal 
frustrated synchronization patterns, but the non-trivial topological effects on the synchronization stability.

The similarities in the graph topological electrical measures between the
power-grids of the two continents, especially the cable length distributions, 
suggest a universality hypothesis, which breaks down as we go towards 
the lower level sub-systems or as the consequence of under-sampling, like subsets of real data. The latter can very well be observed for the obviously incomplete EU22 power-grid case, which provides results, 
resembling to smaller regions, like for the USHW, corresponding to the
North-West states of the US.
In general, universality is expected to occur in the infinite system size
limit, according to statistical physics~\cite{odorbook} and the 
behavior in smaller systems can deviate from it. 
Our extensive, continent sized analysis provides an opportunity to
observe it.

We have also shown a frequency fluctuation analysis of Hungarian data
that can be described by similar super-statistics and $q$ exponent,  
as the ones published for Nordic grid region~\cite{gorjao2021},\cite{gorjao2022}. 
This also strengthens the universality hypothesis we advance in this paper.

\begin{acknowledgments}
Support from the Hungarian National Research, Development and Innovation 
Office NKFIH (K128989) and from the ELKH grant SA-44/2021 is acknowledged.
\end{acknowledgments}

\typeout{} 
\bibliography{hetero}

\end{document}